\documentclass[final,5p,times,twocolumn,authoryear]{elsarticle}

\usepackage{amssymb}
\usepackage[hidelinks]{hyperref}
\usepackage{natbib}
\usepackage{seqsplit}

\usepackage{color}
\usepackage{xcolor}
\usepackage{graphicx}
\usepackage{booktabs}
\usepackage{amsmath, amsthm, amsfonts, amssymb}
\usepackage{mathrsfs}
\usepackage{textcomp}
\usepackage{epstopdf}
\usepackage{multirow}
\usepackage{wrapfig}
\usepackage{subfig}
\usepackage{booktabs} %
\usepackage[ruled]{algorithm2e} %
\usepackage{algorithmicx}  
\usepackage{algpseudocode}
\usepackage{ragged2e}
\usepackage[normalem]{ulem}
\usepackage{cleveref}
\usepackage{bm}

\usepackage{enumitem}

\SetAlFnt{\small}
\SetAlCapFnt{\small}
\SetAlCapNameFnt{\small}
\SetAlCapHSkip{0pt}

\newcommand{\eg}{\textit{e.g.}}

\definecolor{green}{rgb}{0, 0.5, 0}
\definecolor{orange}{rgb}{0.6, 0.3, 0.1}
\definecolor{R-orange}{rgb}{1.0, 0.4, 0.0}
\definecolor{red}{rgb}{1.0, 0.0, 0.0}
\definecolor{teal}{rgb}{0.0, 0.4, 0.4}
\definecolor{purple}{rgb}{0.65,0,0.65}
\definecolor{saffron}{rgb}{0.95,0.75,0.2}
\definecolor{lightblue}{rgb}{0.0,0.5,1}
\definecolor{brown}{rgb}{0.5, 0.16, 0.16}
\definecolor{brickred}{rgb}{.6, .2 .1}
\definecolor{coral}{rgb}{1,0.45,0.33}
\definecolor{newcolor}{rgb}{.8,.349,.1}

\journal{ISPRS Journal of Photogrammetry and Remote Sensing}
\usepackage{lineno}
\begin{document}

\begin{frontmatter}

\title{Building LOD Representation for 3D Urban Scenes}

\author[label1]{Shanshan Pan} %
\author[label1]{Runze Zhang} %
\author[label1]{Yilin Liu} %
\author[label2]{Minglun Gong} %
\author[label1]{Hui Huang\corref{cor}} %
\ead{hhzhiyan@gmail.com}

\cortext[cor]{Corresponding author}

\affiliation[label1]{organization={College of Computer Science and Software Engineering, Shenzhen University},%
            city={Shenzhen},
            country={China}}

\affiliation[label2]{organization={School of Computer Science, University of Guelph},%
            city={Guelph},
            country={Canada}}

\begin{abstract}
    The advances in 3D reconstruction technology, such as photogrammetry and LiDAR scanning, have made it easier to reconstruct accurate and detailed 3D models for urban scenes. Nevertheless, these reconstructed models often contain a large number of geometry primitives, making interactive manipulation and rendering challenging, especially on resource-constrained devices like virtual reality platforms. Therefore, the generation of appropriate levels-of-detail (LOD) representations for these models is crucial. Additionally, automatically reconstructed 3D models tend to suffer from noise and lack semantic information. Dealing with these issues and creating LOD representations that are robust against noise while capturing semantic meaning present significant challenges. In this paper, we propose a novel algorithm to address these challenges. We begin by analyzing the properties of planar primitives detected from
    the input and group these primitives into multiple level sets by
    forming meaningful 3D structures. These level sets form the nodes of our innovative LOD-Tree. By selecting nodes at appropriate depths within the LOD-Tree, different LOD representations can be generated. Experimental results on real and complex urban scenes demonstrate the merits of our approach in generating clean, accurate, and semantically meaningful LOD representations.
\end{abstract}

\onecolumn
\begin{graphicalabstract}
    \vspace{7cm}
    \includegraphics[width=1\linewidth]{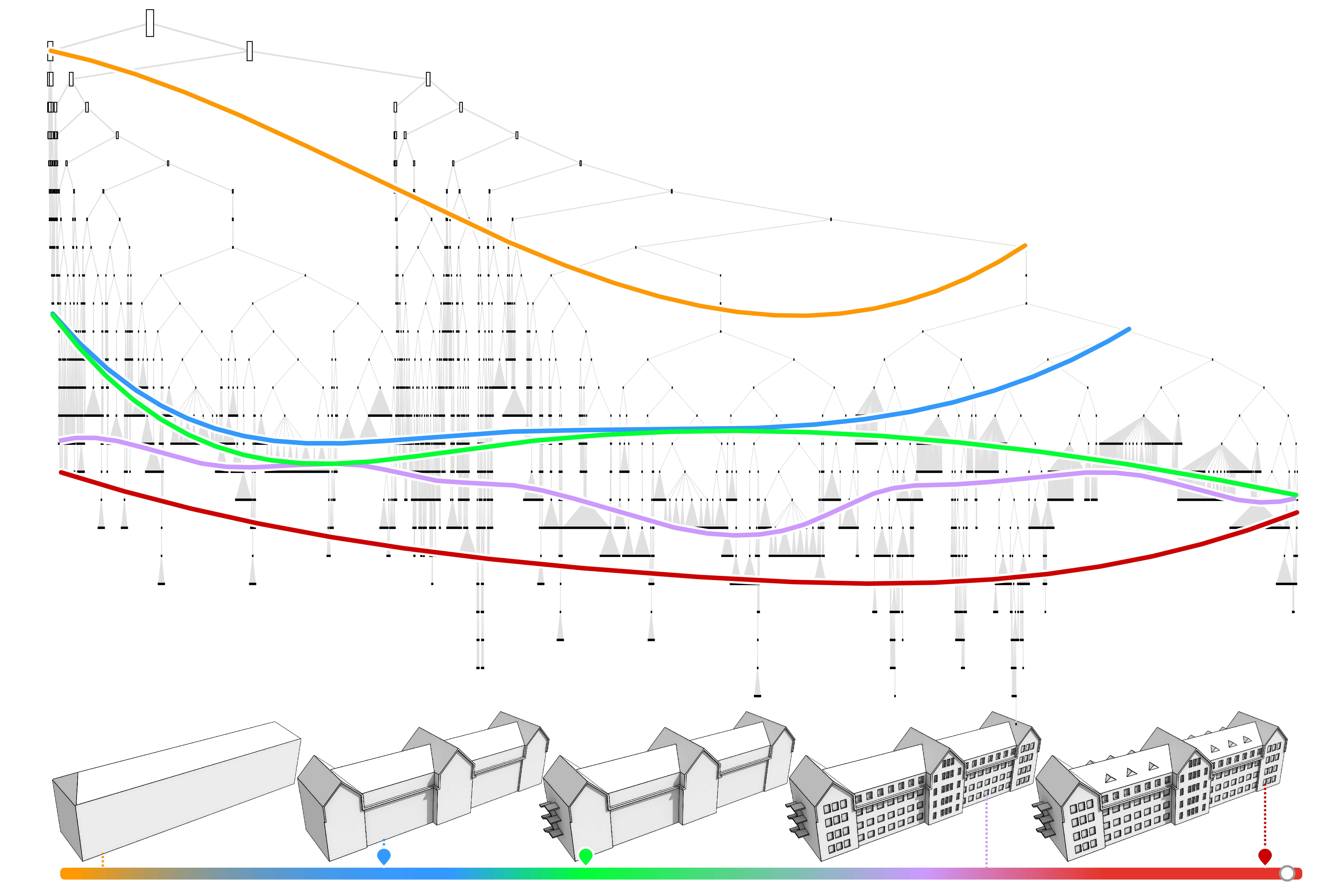}
\end{graphicalabstract}

\begin{highlights}
    \item We propose a novel LOD-Tree representation for flexibly generating LOD models, which is coarse-to-fine and structure-aware.
    \item We introduce a novel Inside/Outside View (IO-View) Analysis for identifying 3D structures.
    \item We present a simple strategy for traversing the LOD-Tree, enabling the generation of models with distinct levels of detail.
\end{highlights}


\end{frontmatter}

\section{Introduction}\label{sec:intro}

The concept of Level-of-Detail (LOD) refers to the usage of multi-resolution models in different scenarios.
It was originally introduced for real-time rendering of complex scenes with limited computational resources~\citep{LOD_origination}, where distant objects can be represented with fewer polygons or simpler textures, whereas objects that are closer to the viewer can be represented with more details. 
Nowadays, LOD models have gone beyond the real-time rendering application and are required by numerous application~\citep{bi2015} such as solar potential estimation~\citep{freitas2015modelling}, studying the thermal characteristics of outdoor spaces~\citep{maragkogiannis2014combining}, firefighting simulations~\citep{chen2014application}, and multi-scale navigation~\citep{hildebrandt2014assisting}. Each of these applications demands specific requirements for the LOD models, highlighting the need for advanced LOD representations to effectively manipulate and render 3D urban scenes.

Current LOD generation methods can be classified into two categories. Visual-dependent methods~\citep{low-poly, neuralLOD, QEM, hoppe1996progressive} use \textit{low-level geometric cues} to iteratively simplify the model to obtain a multi-scale representation. The simplification process usually depends on geometric or visual measurements. These methods offer great flexibility as users can navigate to any intermediate state to obtain the desired model, ensuring smooth visual transitions between models. However, this flexibility can result in numerous semantically meaningless intermediate states, posing challenges for further modeling or analysis.

In contrast, semantic-aware methods~\citep{verdie2015lod, nan2015template, BigSUR, zhu2018large, han2021urban} adhere to predefined LOD definitions. 
To generate LOD models that meet these definitions, these methods require precise identification and segmentation of structures within the input data. However, due to the current lack of large-scale annotated urban datasets for learning high-level features, existing methods resort to using geometric properties for semantic-aware segmentation. This reliance on fixed LOD definitions not only limits their ability to represent complex geometries but also inevitably leads to ambiguities ~\citep{improvedGML} and loss of details ~\citep{verdie2015lod}.

In this paper, we introduce a novel \emph{LOD-Tree} representation for flexibly constructing concise and structural LOD models for 3D urban scenes.
To build the LOD-Tree, we first propose \emph{Inside/Outside View (IO-View)} to analyze the properties of planar primitives detected from the input and group these primitives into multiple level sets by forming meaningful 3D structures. Unlike previous semantic-aware methods~\citep{verdie2015lod, nan2015template}, our IO-View analysis algorithm considers not only the geometric attributes of individual planar primitive but also their interrelationships. Subsequently, these level sets form the nodes of our innovative LOD-Tree. Nodes corresponding to principal primitives, such as roofs and walls, are positioned closer to the root, while nodes containing secondary primitives, such as doors and windows, are closer to the leaves. Finally, we propose a strategy for traversing the LOD-Tree, starting from the root and combining nodes at different depths to flexibly extract models of different levels of detail.

Our contributions can be summarized as follows:
\begin{itemize}[leftmargin=*]
    \item A novel LOD-Tree representation for flexibly generating LOD models, which is coarse-to-fine and structure-aware.
    \item Introducing IO-View, a novel plane analysis method for identifying 3D structures.
    \item A method for traversing the LOD-Tree to identify models with distinct LOD variations.
\end{itemize}

We validate our method on 21 real-world datasets to demonstrate our ability to extract structure-aware LOD models. 
Our comparisons include evaluations against the vanilla BSP-Tree, state-of-the-art visual-dependent and semantic-aware LOD approaches, and even human modeling.
Compared with other methods, we reach a better semantic-geometric trade-off, thus producing more concise and high-quality LOD models without relying on heuristic assumptions about semantic relations.
Additionally, we conduct a detailed analysis of various parameters, intermediate outcomes, and exceptional scenarios that impact the LOD-Tree, thus providing substantial evidence of its resilience.
Notably, we highlight that the generation of LOD models is user-interactable while maintaining the simplicity and applicability of the generated results.

\section{Related Work}
\label{sec:rw}
Level-of-detail (LOD) generation can be seen as a particular type of structural reconstruction where different metrics have been proposed to distinguish the level of the model.
In Sec.~\ref{sec:rw_lod}, we review two common approaches in LOD generation, \textit{Visual-dependent LOD} and \textit{Semantic-aware LOD}.
We also discuss the vanilla structural reconstruction in Sec.~\ref{sec:rw_structural}, as it usually plays an important part in the current LOD generation methods. The related interactive approaches are briefly reviewed in Sec.~\ref{sec:rw_interactive}.
 
\subsection{LOD Generation}
\label{sec:rw_lod}
\paragraph{\textbf{Visual-dependent LOD}} 

A common approach is based on the visual effect when the model moves away from the viewer to determine which LOD model shall be used.
As the model moves further, the reduced visual quality of the model is often unnoticed.
Thus, a coarser model can be used for distant geometry.
Visual-dependent LOD usually focuses on the geometry complexity of the models, that is, simplifying the model 
while maintaining the object's visual appearance as much as possible.

General mesh simplification or approximation methods~\citep{QEM, hoppe1996progressive, lindstrom2000out} iteratively collapse the local surface patch until the model reaches a target number of faces. To better preserve the planar structure of objects, 
the edge contraction operators impose a planarity constraint~\citep{salinas2015structure, li2021feature}. 
Even so, adjacent faces of the resulting triangular mesh are unlikely to be co-planar.
These methods are generally efficient but strict about the input shape and quality. It is challenging to preserve the delicate structural features of objects, such as the roof boundaries. 

More advanced approaches~\citep{low-poly,robust-lowpoly,takikawaVariableBitrateNeural2022, lindstrom2000image} take into account the pixel-wise metric, 
aiming to preserve the visual similarity of the input.
For example, \citet{low-poly} first preserves silhouettes that are important for visual appearance and then enriches the details by progressively carving the mesh. 
\citet{robust-lowpoly} further extends this idea by introducing a mixed remeshing and simplification framework, which can handle more complex inputs.

Some classic approaches~\citep{frisken2000adaptively, lorensen1987marching} use octrees to discretize the Euclidean space. The geometric LODs can then be determined by the resolution or tree depth, 
and different LODs can be blended with interpolation.
Inspired by this, \citet{neuralLOD} discretizes the space using a sparse voxel octree (SVO) that holds a collection of learned features. 
It enables continuous LOD with geometric-aware SDF interpolation and truncates the tree depth while inheriting the advantages of classic approaches.
They further compress this partition using a learned dictionary~\citep{takikawaVariableBitrateNeural2022}, which reduces memory consumption without losing accuracy.

Unfortunately, most of these methods rely on geometric cues, and the ignored semantics and topology structures usually induce a large amount of meaningless intermediate states.

\paragraph{\textbf{Semantic-aware LOD}} 
An alternative definition of LOD comes from cityGML ~\citep{cityGML}.
It is intended to differentiate models at different LODs based on their high-level semantic properties.  

Existing methods~\citep{verdie2015lod, nan2015template, BigSUR, zhu2018large, han2021urban} 
mainly reconstruct the LOD-2 model first, that is, a model with a simplified roof shape.
\citet{verdie2015lod} detect the planes belonging to the roof and facade by the size of the area and then generate a simplified LOD-2 model by a plane-based structural reconstruction method. 
\citet{zhu2018large} and \citet{han2021urban} only reconstruct the roofs and 
then generate the model by lifting the roof boundary polygons to the roof planes.
\citet{BigSUR} fuse building footprints with vertical profiles and the attached building facades to produce the structural model.
After generating the LOD-2 model, the LOD-3 model is usually generated by collecting the details of facades and roofs from the images. 
LOD-1 can be obtained by calculating the average height of LOD-2, and LOD-0 is the 2D footprint of the model. Recently, Scan2LoD3~\citep{wysocki2023scan2lod3} introduced a unified framework that leverages multimodal data (e.g., images or pre-defined LOD-2 models) and Bayesian fusion to enhance LOD-3 model construction.
Commonly, they all focus on generating coarse models that are only approximated with a small number of textured planes and require additional image information to localize details that are mostly in 2D representations.
Although \citet{nan2015template} can incorporate geometric details into the coarse models through template assembly, they can only enrich details depending on the information provided by facade images, which limits its applicability in diverse scenarios.

On the other hand, the fixed LOD definition also restricts its ability to represent a complex scene. \citet{improvedGML} extend the definition of LOD from 5 states to 16 states by enumerating the combination of the common items, \eg, window, door, and chimney. However, it is still unclear how to represent and modify unseen items in real cases that are very diverse.
In contrast, our approach addresses broader scenarios by relying solely on MVS point clouds or dense reconstructions, eliminating the need for supplementary data such as images or pre-defined LOD-2 models. Our method can directly extract diverse detailed structures, including chimneys, dormers, and air-conditioning units. This not only surpasses the structural granularity of conventional LOD-2 and LOD-3 classifications but also enables flexible adjustment of detail levels to meet various downstream applications, such as solar panel installations. By doing so, our approach provides greater adaptability and versatility for urban modeling in real-world scenarios.

\begin{figure*}[t]
	\centering
	\includegraphics[width=1\textwidth]{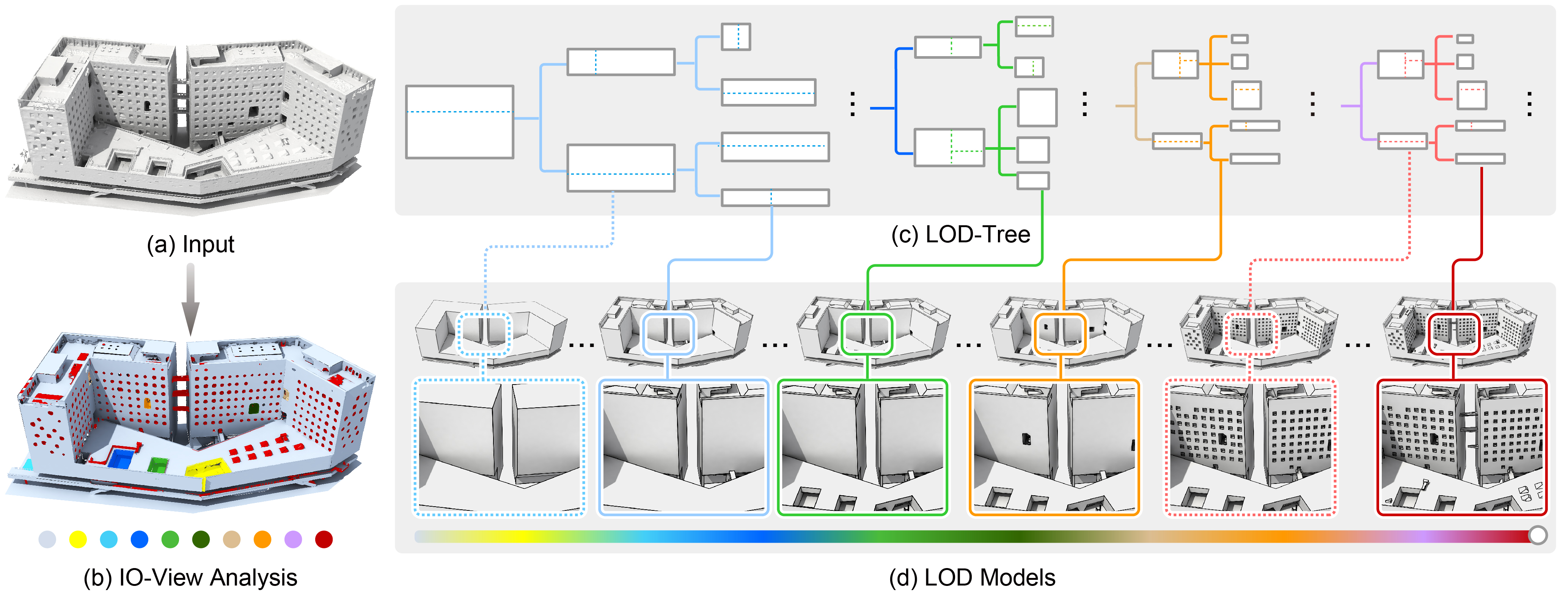}
	\caption{
      Overview: Our method takes a dense triangle mesh or point cloud as input (a) and performs an IO-View analysis to analyze the properties of planar primitives. This analysis automatically groups these planar primitives into multiple level sets, as shown in different colors in (b). We then sequentially partition the space using these level sets to generate a LOD-Tree (c). Each selected combination of tree nodes corresponds to an output model. We showcase six representative models (d) derived from the LOD-Tree. 
	  }
	\label{fig:overview}
\end{figure*}

\subsection{Structural Reconstruction}
\label{sec:rw_structural}
Structural reconstruction consists of extracting geometric primitives from point clouds or a dense mesh and then assembling them into a compact parametric 3D model.
 
The common approaches~\citep{chen2022DIF, KSR, polyfit, Chauve}
operate by partitioning the 3D space of the input using detected planar primitives. 
Thus the 3D space is represented as a decomposition of convex polyhedral cells.
The output mesh is then extracted by labelling the cells as the inside or outside of the surface.
Here, how to partition the space is a core, based on which existing approaches can be classified into three groups.
\citet{polyfit} split the space exhaustively by intersecting each pair of planes. 
Observing that primitives mostly only affect the local space, \citet{KSR} use the \emph{grow-and-collide} scheme to generate a smaller number of meaningful cells.
Considering the time complexity, \citet{chen2022DIF} propose an adaptive space partition strategy performed in a hierarchical manner. The 3D space is recursively divided to build a binary space partitioning tree (BSP-Tree). However, this leads to numerous meaningless subspaces. To tackle this, \citet{sulzer2024concise} reorders the plane splitting process to ensure each binary partition is as effective as possible, enabling the construction of a more compact BSP tree with fewer splits and subspaces.

On the other hand, researchers also try to detect planar primitives in different scales, which can be employed to construct structural LOD models. 
\citet{FangPlanarShape} extract high-level representations using planar primitives at different scales. 
Starting from the finer scale, the planar primitives are progressively merged to yield primitives at coarser levels. 
Similarly, \citet{LejemblePersistence} generate a graph whose nodes represent detected planar primitives in different scales. 
The principal planes at various scales can be efficiently extracted by examining the nodes' connections.
Other advanced planar detection and regularization methods~\citep{YuL22,Mitra15} can be used to improve the detection results. 
However, most techniques merely offer the low-level geometric characteristics of the planar primitives (\eg, their areas), while ignoring the relationship between planes and the effect of plane order on space partition.
\citet{sulzer2024concise} and our method share a similar spirit in aiming to construct a more compact and semantically aware BSP tree. \citet{sulzer2024concise} adopts a greedy strategy to redefine the order of space partitioning, whereas our method leverages a global analysis of inter-planar relationships to guide the partitioning process.

\subsection{Interactive Modeling}
\label{sec:rw_interactive}
Commercial software such as 3ds Max and SketchUp gives users a great degree of freedom to operate.   
Taking input as a reference, professional artists gradually generate the required clean models 
by operating basic primitives such as points, lines, planes and boxes. 
Since these advanced software are designed for domain experts, they are not friendly to novice users.

For more intuitive interaction, one possible solution called procedural modelling consists of encoding and generating facade layouts using synthetic rules. The downside of procedural modelling is that it needs expert specifications of the rules and is limited in the realism of the resulting models and their variations. Furthermore, it is complicated to formulate the rules to construct the existing objects, such as buildings, exactly.

Another popular solution of interaction is based on optimization~\citep{O-Snap, Smartboxes, ren2021roof}.
\citet{O-Snap} allow the user to modify the sketches with coarse and loose 2D strokes, as
the exact alignment of the polygons is automatically performed by snapping.
\citet{Smartboxes} manipulate simple building blocks to assemble detailed 3D primitives and utilize the regularity of facades to balance between data fitting and structural regularity terms. 
\citet{ren2021roof} propose an interactive roof editing framework, 
which can be used for roof design or reconstruction from the captured aerial images. 

All these works tend to focus on specific reconstructions of buildings. As far as we know, there is currently no convenient interactive solution for LOD generation. Therefore, we propose a handy tool (LOD-Tree) that allows users to efficiently select the desired structural LOD models with a simple sliding operation; see our accompanying demonstration video.

\section{Problem Statement and Overview}\label{sec:overview}

\paragraph{\textbf{Problem Statement}} 

  Given an oriented triangle mesh or a point cloud, we aim to build a LOD-Tree, which is then used to generate semantic-aware LOD models.
  To better facilitate LOD generation, our target LOD-Tree representation should satisfy the following properties:
 
  \begin{itemize}[leftmargin=*]
    \item Structures become progressively enriched with the traversal of the tree.
    \item Principal structures should appear earlier, and secondary structures later. Specifically, nodes corresponding to principal primitives (e.g., roofs and walls) are closer to the root, while nodes containing secondary primitives (e.g., doors and windows) are nearer to the leaves.
    \item All LODs generated by LOD-Tree are represented by concise and watertight polygonal meshes.
  \end{itemize}

\paragraph{\textbf{Method Overview}}
 To build a LOD-Tree, we initially detect planar primitives $P = \{p_i\}_{i=0}^N$ from the input using region growing~\citep{RegionGrowing}. Then, the IO-View technique we proposed is used to analyze the properties of the primitives and group them into multiple level sets by forming meaningful 3D structures. These level sets guide the space partitioning process, transforming traditional binary space partitioning into a semantic-aware LOD-Tree representation. Finally, we design a simple traversal strategy to extract LOD models from the LOD-Tree. Our framework is shown in Fig.~\ref{fig:overview}.

\section{Methodology}\label{sec:method}

\subsection{IO-View Analysis}\label{sec:method_IOView}

Given an input model $I$, an initial set of planar primitives $P = \{p_i\}_{i=0}^N$ is extracted by region growing~\citep{RegionGrowing} (default detection parameters: $\epsilon$ = 0.15m, $\theta$ = 40$^\circ$, $\sigma$ = 15). Our proposed \emph{IO-View} aims to group and sort these planar primitives by forming enclosed 3D structures. In this section, we first discuss the initialization of the IO-View process. We then elaborate on the extraction and grouping of 3D structures. These structures are used to sort the planar primitives. Additionally, we introduce an adaptive regularization process designed to enhance the quality of these planar primitives.

\paragraph*{\textbf{Initialization}}

Given the initial set of planar primitives $P = \{p_i\}_{i=0}^N$, we can obtain a set of polyhedra by binary space partitioning in existing structural reconstruction methods~\citep{chen2022DIF}. Also, each polyhedron contains an in/out label based on a ray-casting strategy~\citep{raystabbing}, indicating whether the polyhedron is located inside or outside of the model $I$. 
The details of the binary space partitioning and polyhedra in/out labelling are shown in ~\ref{sec:bsp} and ~\ref{sec:labelling}, respectively.

\paragraph*{\textbf{Analysis: extract 3D structures}}  

\begin{figure}[t!]
	\centering
	\includegraphics[width=0.9\linewidth]{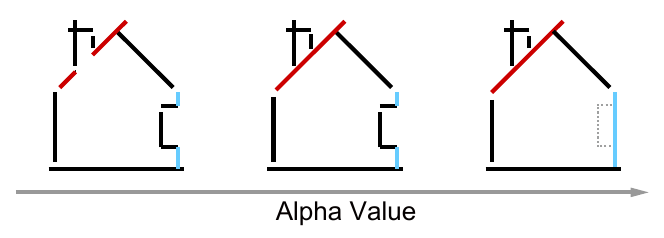}
	\caption{
		The effect of the $\alpha$-value on the $\alpha$-shape. As the $\alpha$-value goes higher, the hole on the planar primitives can be filled. The Top row shows the boundary changes of a model composed of multiple planes, while the bottom row shows the boundary changes of a single plane.
	}
	\label{fig:alpha-value}
\end{figure}

To group planar primitives, we first need to identify which 3D structures these primitives constitute in the input.
We achieve this by first distinguishing between the principal and secondary structures. We observe that architectural details are often connected to principal surfaces. Balconies and chimneys are examples of addon volumes to the overall model, while windows and doors are commonly represented as cutout volumes. These secondary structures create holes on the principal surfaces where they attach. To separate these secondary structures and the principal structure, we simply use a set of $\alpha$-shapes~\citep{alpha_shape} to represent the detected planar primitives. The property that the $\alpha$-shape changes with the $\alpha$-value allows us to fill these holes by adjusting the $\alpha$-value. These $\alpha$-shapes are essential for extracting 3D structures. In our experiments, we typically set the $\alpha$-value to 7m; see Fig.~\ref{fig:alpha-value}.

\begin{figure}[t!]
	\centering
	\includegraphics[width=1\linewidth]{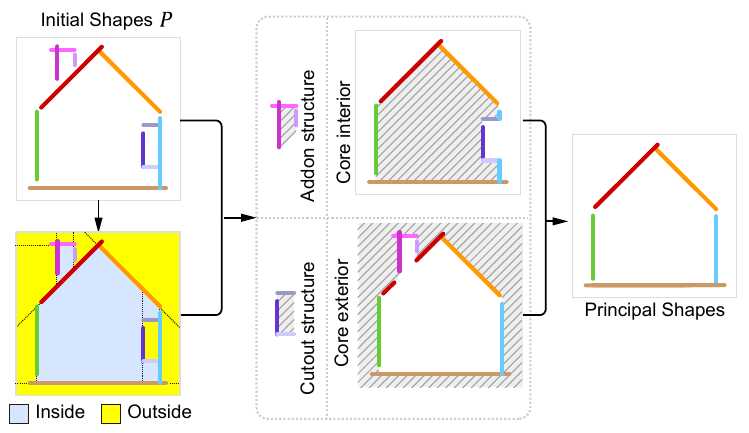}
	\caption{
		We first detect multiple $\alpha$-shapes. Each colored line corresponds to an $\alpha$-shape; as shown in the upper left. 
		These $\alpha$-shapes split the space into polyhedra, and two adjacent polyhedra share a common face; as shown in the bottom left.  
		By removing faces that are not covered by an $\alpha$-shape, 
		we can divide the entire space into four categories, as shown in the middle. 
		Finally, we extract the primitives shared by ``core interior'' and ``core exterior'' as the principal primitives $S_{0}$; see the colored lines on the right figure.
		The remaining planar primitives are labelled as secondary primitives, which define addon (3 pinkish lines) and cutout (3 blueish lines) structures.
	}
	\label{fig:io-view}
\end{figure}

Specifically, since the partitioning of planar primitives forms the polyhedra at the initialization process, it is straightforward to know whether the faces of the polyhedra is covered by $\alpha$-shapes; see the left side of Fig.~\ref{fig:io-view}. Utilizing the corresponding between $\alpha$-shapes and polyhedra, we can efficiently extract the 3D structures. If adjacent polyhedra have the same in/out labels and are not separated by any $\alpha$-shape, we iteratively merge them. Once all polyhedra are merged into multiple \emph{regions}, \eg, the middle in Fig.~\ref{fig:io-view}, we define these regions using the following four categories:
\begin{itemize}[leftmargin=*]
	\item[$\bullet$]Core interior $V_{in}$: the region that has the largest overlap with the inside of input model $I$; 
	\item[$\bullet$]Core exterior $V_{out}$: the region that has the largest overlap with the outside of input model $I$;
	\item[$\bullet$]Addon structures $\{V_{+}\}$: other regions inside input model $I$;
	\item[$\bullet$]Cutout structures $\{V_{-}\}$: other regions outside input model $I$.
\end{itemize}

Finally, $\{V_{+}\}$ and $\{V_{-}\}$ constitute the secondary structures of the building, while the planar primitives that separate regions $V_{in}$ and $V_{out}$ form the principal structure, as illustrated on the right side of Fig.~\ref{fig:io-view}.

\paragraph*{\textbf{Analysis: group 3D structures}} 

Based on the secondary structures (i.e., addon and cutout structures), we further perform a \emph{two-stage clustering} to aggregate similar structures.

In the first stage, all secondary structures are grouped into clusters $C=\{C_i\}_{i=0}^{N_c}$. Structures within each cluster $C_i$ are considered to have similar semantic meanings. Specifically, we iterate each primitive $p$ of the principal structure, calculate the projected area of the secondary structures located on $p$, and perform the mean shift clustering (bandwidth = 2) based on the projected area. 

We further perform the second-stage mean shift clustering (bandwidth = 4) on all clusters $C$ to obtain multiple level sets $L=\{L_i\}_{i=1}^{N_l}$, based on the average volume size of each cluster $C_i$. This is because some clusters usually have similar importance to visual appearance or practical applications, \eg, doors and windows are classified into one level in CityGML.

\begin{figure*}[t]
	\centering
	\includegraphics[width=1\textwidth]{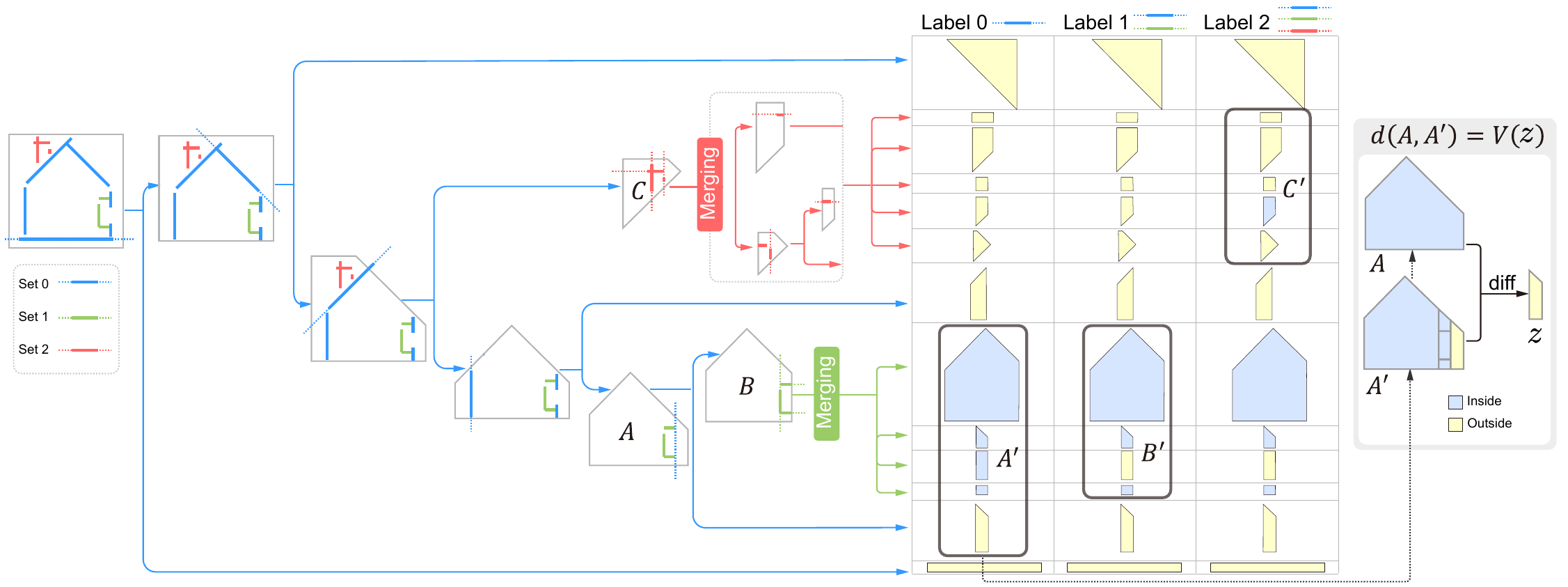}
	\caption{
		Given three level sets $\{S_0, S_1, S_2\}$ (blue, green and red), we gradually get the LOD-Tree divided by these three sets and generate multiple in/out labels (Label 0, Label 1, Label 2) for leaf nodes to facilitate the construction of LOD-Tree.
		The label $k$ of the leaf nodes is determined by a majority vote based on level sets $\bigcup_{i=0}^k S_i$. 
		The diff-value of each node depends on the similarity of the space associated with this node and all its leaf nodes by comparing their labels and sizes. 
		That is, node $A$ is labelled as $in$, while leaf nodes of $A$ are labelled as $A'$ (circled in black), then we compute the difference of $A$ and $A'$ as the diff-value of $A$.
		A node with a higher diff-value is considered less similar to the finest model and, therefore, more valuable when splitting the space. In particular, if a cluster contains fewer than 10 primitives per structure, we will perform a merging operation to merge BSP nodes that are split by the cluster's primitives into a single LOD-node (\eg, windows, chimneys). That is, when we reach node $C$, if we want to cut it next time, node $C$ will be expanded into five leaf nodes at one time. For simplicity, multiple windows are represented by a single window in the diagram, and the same goes for chimneys.
	}
	\label{fig:label_and_cut_value}
\end{figure*}

\paragraph*{\textbf{Regularization}}

Based on the principal structure and grouped secondary structures, we proceed with the regularization~\citep{verdie2015lod} of the planar primitives to enhance the quality of the LOD models.
We observe that the scan for detailed geometry often exhibits lower quality than the principal structure in real-world cases. This is primarily attributed to the limited number of samples acquired for small structures. To address this, we apply stronger regularization techniques specifically tailored for primitives of secondary structures. The procedure is outlined as follows:

\begin{itemize}[leftmargin=*]
	\item[$\bullet$] For primitives of the principal structure, we perform mild regularization of parallelism, orthogonality and coplanarity (angle = 5$^\circ$, distance = 0.01m). 
	\item[$\bullet$] For clusters of the secondary structures, we take the principal primitive $p$ where the cluster is located as the reference primitive and perform stronger regularization when the cluster’s primitives are approximately parallel or perpendicular to $p$ (angle = 15$^\circ$, distance = 0.01m). In particular, if $p$ is roughly perpendicular to the XY plane (within a 5$^\circ$ angle threshold) and the cluster contains more than three cutout structures, these structures are identified as windows and replaced with template meshes in the form of cuboids. 
\end{itemize}

\paragraph*{\textbf{Sort the planar primitives}}

Once regularization is finalized, we sort the planar primitives. We first sort $L$ based on the average volume of each level set. Next, we sort clusters in each level set $L_i$ according to the average volume of the clusters. Then, within each cluster $C_i$, we rank the primitives by their area size. Finally, we obtain the scale-sorted planar primitives $S=\{S_i\}_{i=0}^{N_l}$. Here, $S_{0}$ represents primitives of the principal structure, while $\{S_{1},...,S_{N_l}\}$ correspond to the primitives of the secondary structures at each level.

\subsection{LOD-Tree Generation}\label{sec:method_LODTree}

Building upon the scale-sorted planar primitives $S = \{S_i\}_{i=0}^{N_l}$, we proceed to construct the LOD-Tree as a novel representation for levels of detail.

The LOD-Tree is built upon a BSP-Tree, where each node corresponds to a specific space. Therefore, we first discuss the generation of the BSP-Tree using scale-sorted primitives. Subsequently, we demonstrate how to merge BSP nodes into LOD-nodes and construct the LOD-Tree. Finally, we present the process of extracting LOD models by traversing the LOD-Tree.

\paragraph*{\textbf{Initialization: generate a new BSP-Tree}}

We use the scale-sorted primitives $S = \{S_i\}_{i=0}^{N_l}$ to split the space and generate the BSP-Tree. 
The BSP-Tree construction process begins with the recursive division of the bounding box of the model. Each plane partitions a space into two subspaces. During the operation, unprocessed primitives are assigned to their corresponding subspace. If a primitive spans both subspaces, it is split into two to ensure that each primitive in the new subspace remains within its designated region. This partitioning process continues until there are no more subspaces that can be divided. At the end of this process, each leaf node in the BSP-Tree corresponds to a convex polyhedron cell, and all the leaf nodes are combined to form the initial bounding box of the model.

\paragraph*{\textbf{Merging: merge correlated BSP nodes}}
Once the BSP-Tree is constructed, we further merge correlated BSP nodes into a single LOD-node, allowing this LOD-node to be partitioned by multiple primitives simultaneously. 
Specifically, since structures composed of a small number of primitives offer limited interpolation space, if the cluster contains less than $K$ primitives per structure, we merge BSP nodes that are split by the cluster's primitives into a single LOD-node. In the experiments, we set $K$ to 10 by default.
The child nodes of the merged BSP nodes become the children of the newly formed LOD-node. This process continues until all BSP nodes have been processed, as illustrated in Fig.~\ref{fig:label_and_cut_value}. This results in a more compact and structural LOD-Tree representation.

LOD-Tree has two improvements over BSP-Tree. Firstly, LOD-Tree allows nodes to be partitioned by multiple primitives at the same time, which can quickly and completely add some structures to the LOD model simultaneously. However, BSP-node is partitioned by an individual primitive which is meaningless most of the time; for example, a window structure composed of five primitives might require five separate partitions to carve out the original shape. Moreover, by leveraging our approach to 3D structure recognition, our partitioning sequence ensures that larger structures are prioritized over smaller ones. In contrast, traditional BSP-Tree relies on plane area size for partitioning, which cannot necessarily reflect the importance of the structures.

\paragraph*{\textbf{LOD-Tree: traverse the tree using diff-value}} 

Finally, we propose a strategy for traversing the LOD-Tree, starting from the root and combining nodes at different depths to flexibly extract models of different levels of detail.
To guide the selection of LOD-nodes, we assign a \emph{diff-value} to each node.
This diff-value is computed as the absolute volume difference between the 3D shape modelled by node $n$ and its corresponding leaf nodes.
Nodes with higher diff-values indicate a higher approximation error and thus provide more potential for improvement.

To calculate the diff-value for each node, we first determine the in/out labels of the nodes. For each leaf node, multiple labels are generated based on the level sets, where label $k$ is determined by the level sets $\bigcup_{i=0}^k S_i$ (also based on the ray-casting strategy). The in/out label of a non-leaf node is determined by the label of its leaf nodes. Specifically, we first identify which level's primitives partition the node $n$. 
At this level, the node $n$ is labelled as $in$ if the total volume of its leaf nodes that are labeled $in$ exceeds 65\%.
Once the labelling is determined, we compute the diff-value $d(n, LN(n))$ of each non-leaf node $n$ which is split by the primitives at level $k$. The diff-value is computed as follows:

$$ d(n, LN(n)) = abs\left( x_k(n)  V(n)-\sum_{m \in LN(n)}{x_k(m)V(m)} \right),$$
where $LN(n)$ are the leaf nodes of node $n$ on the whole tree, 
$x_k(\cdot)$ is a binary function that indicates whether a node is labelled as $in$. $V(\cdot)$ represents the volume of the node space; see Fig.~\ref{fig:label_and_cut_value}.

The tree traversal is implemented using a priority queue $Q$ that stores combinations of nodes. This queue is populated by sorting the diff-values of these nodes in ascending order. Starting from the root, we greedily select the LOD-node with the maximum diff-value from $Q$ as the next target for splitting. We then update $Q$ by removing the selected node and pushing its children into the queue.

\begin{algorithm}[t!]
	\caption{Traverse the tree using diff-value}\label{algo:algo2}
	\SetKwInOut{Input}{input}\SetKwInOut{Output}{output}
	\SetKwComment{Comment}{//}{}
	\SetKwComment{Com}{$\triangleright$ }{}
	\Input{LOD-Tree $T$}
	\Output{candidate LOD model set $M$} 
	$M \leftarrow \emptyset$\;
	$Q \leftarrow T.root$\;
	$level \leftarrow 0$\;
	\While{$level < N_l$}
	{
		\eIf{$DiffSum(Q)==0$}
		{
			\Com{Arrive at an anchor}
			\Com{Turn to the next level}
			$level \leftarrow level+1$\;
			$M \leftarrow M \cup extract(Q)$\;
			$Continue$\;
		}
		{
			$find\ best\ node\ n\ in\ Q\ to\ expand$\;
			$Q.erase(n)$\;
            $Q.add(n.children)$\;
			\If{$is\_interpolation(Q)$}
			{
				$M \gets M \cup extract(Q)$\;
			}
		}
	}
\end{algorithm}

\begin{figure}[t!]
	\centering
	\includegraphics[width=1\linewidth]{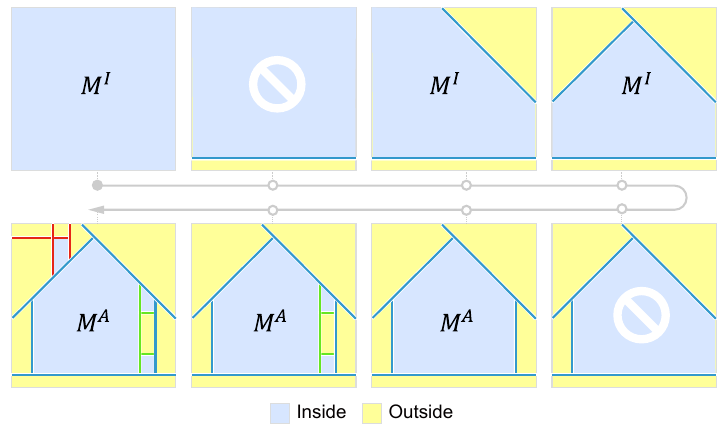}
	\caption{
		Illustration of the evolution of priority queue $Q$ and LOD model extraction. Interpolation models ($M^I$) are extracted when the appearance changes significantly. Anchor models ($M^A$) are extracted when each level set $S_i$ completes space partitioning. Both model types are stored in the candidate LOD model set $M$.
	}
	\label{fig:lodmodels}
\end{figure}

The sum of the diff-values in $Q$ reflects the total approximation error between the current model and the most detailed model achievable under the current level. As a result, as we traverse deeper into the LOD-Tree, the corresponding model transitions smoothly. However, we observe that not all combinations of nodes in Q are meaningful. To avoid this situation, we only extract models when the summed diff-value drops significantly. Specifically, if the summed diff-value drops to below a certain percentage of the previous model, we extract a new model. We set this percentage to 80\% by default. The models extracted through this rule are referred to as \emph{interpolation models} ($M^I$).
When the summed diff-value of $Q$ reaches zero, the corresponding model represents the most detailed representation captured under the current level. Such models are referred to as \emph{anchor models} ($M^A$). Both anchor and interpolation models are extracted automatically and stored in the candidate LOD model set $M$; see Algorithm~\ref{algo:algo2} and Fig.~\ref{fig:lodmodels}. Users can then select all or part of the models in $M$ as LOD output.

\paragraph*{\textbf{Interactive Phase}} 
Since the effect of plane detection limits the IO-View analysis, the generated anchor models cannot guarantee a perfect solution in all cases.
Therefore, we generate interpolation models automatically to balance the completeness and low redundancy of sampling from LOD-Tree, providing more possibilities. 
We provide a simple interactive interface. Users can select the appropriate LOD model from the candidate LOD model set $M$ based on their specific needs. For example, a solar panel installation may require detailed roof information while ignoring facade details. 
Note that the number of our candidate LOD models typically ranges between 10 and 70.

\section{Experimental Results}\label{sec:results}

\paragraph{\textbf{Datasets}}
To evaluate the effectiveness of our proposed methodology, we conducted experiments on a diverse range of real-world scenes with different architectural styles and functions. Our dataset consists of 21 models obtained from various sources such as public datasets (\eg, Polytech~\citep{lyl-2022}, Hitech~\citep{DroneScan20}, Residency~\citep{zhang2021continuous}, Church and Lans~\citep{KSR}) and models captured using a single-camera drone (\eg, AviTech, School, Hall, CSSE, and Factory). These models exhibit complex geometry and topology, with many of them being non-manifold and composed of multiple components. Table~\ref{table:input} provides statistics of input models in our dataset, and presents the running time of each stage in our algorithm. The Church, Lans, and House models are point cloud data, so the number of faces, components and non-manifold vertices cannot be quantified.

\begin{table*}[t!]
    \newcommand{\tabincell}[2]{\begin{tabular}{@{}#1@{}}#2\end{tabular}}
    \caption{
        Columns 2 to 6 present statistics for input models, including the number of faces ($F$), vertices ($V$), components ($C$) and non-manifold vertices ($M$), as well as the length of the bounding box diagonal ($D$).
        Columns 7 to 11 showcase the performances of our algorithm on various models, detailing the number of initial planar primitives $P$, the number of scale-sorted planar primitives $S$, as well as the runtime for IO-View analysis $T_{1}$(Sec.~\ref{sec:method_IOView}), LOD-Tree construction $T_{2}$(Sec.~\ref{sec:method_LODTree}), and LOD-Tree traversal $T_{3}$ (Sec.~\ref{sec:method_LODTree}), respectively.
    }
    \centering
    \label{table:input}
    \begin{tabular}{cccccccccccccc}
        \toprule
        & Models      & $F (\#)$   & $V (\#)$   & $C (\#)$    & $M (\#)$  & $D (m)$   & $P (\#)$  & $S (\#)$  & $T_{1} (s)$  & $T_{2} (s)$  & $T_{3} (s)$ \\ \hline 
        
        & Polytech (Fig.~\ref{fig:overview})         & 1984k     & 1362k      & 23170       & 236    & 182          & 6154  & 4525  & 60    & 92    & 80        \\
        & AviTech (Fig.~\ref{fig:Reg})               & 723k      & 455k       & 5836        & 178    & 132          & 4637  & 1324  & 27    & 24    & 12        \\
        & School (Fig.~\ref{fig:gallary})            & 1456k     & 998k       & 22714       & 724    & 170          & 9216  & 2145  & 66    & 50    & 24        \\
        & Hall (Fig.~\ref{fig:gallary})              & 587k      & 525k       & 31694       & 333    & 104          & 4181  & 2384  & 28    & 59    & 34        \\
        & CSSE (Fig.~\ref{fig:gallary})              & 1724k     & 1363k      & 49125       & 747    & 185          & 13245 & 4799  & 110   & 102   & 135       \\
        & Church (Fig.~\ref{fig:gallary})            & -         & 211k       & -           & -      & 64           & 850   & 290   & 3     & 2     & 1         \\
        & Lans (Fig.~\ref{fig:gallary})              & -         & 1220k      & -           & -      & 48           & 1950  & 226   & 10    & 2     & 1         \\
        & House (Fig.~\ref{fig:failure})             & -         & 371k       & -           & -      & 46           & 1181  & 463   & 9     & 5     & 1         \\
        & Residency (Fig.~\ref{fig:residence})       & 9307k     & 4693k      & 370         & 8      & 556          & 1972  & 1257  & 159   & 77    & 56        \\ 
        & Factory (Fig.~\ref{fig:vs_structural})     & 136k      & 102k       & 2922        & 57     & 102          & 425   & 130   & 3     & 2     & 1         \\
        & Cottage (Fig.~\ref{fig:QEM})               & 150k      & 76k        & 2           & 0      & 68           & 1033  & 574   & 6     & 10    & 3         \\
        & Headquarter (Fig.~\ref{fig:vs_low_poly})   & 1217k     & 768k       & 10392       & 268    & 153          & 8149  & 1382  & 51    & 39    & 17        \\
        & Department (Fig.~\ref{fig:vs_low_poly})    & 574k      & 405k       & 11010       & 249    & 174          & 5810  & 1532  & 37    & 30    & 20        \\
        & Hitech (Fig.~\ref{fig:vs_low_poly})        & 1936k     & 968k       & 26          & 1      & 178          & 4403  & 1469  & 51    & 27    & 31        \\ 
        & Mall  (Fig.~\ref{fig:vs_low_poly})         & 388k      & 236k       & 12          & 0      & 134          & 3065  & 1001  & 18    & 17    & 5         \\ 
        & Office (Fig.~\ref{fig:vs_low_poly})        & 91k       & 47k        & 12          & 5      & 76           & 659   & 297   & 3     & 4     & 2         \\
        & Hotel (Fig.~\ref{fig:vs_low_poly})         & 601k      & 302k       & 7           & 0      & 79           & 3089  & 949   & 18    & 16    & 4         \\
        & Bank (Fig.~\ref{fig:vs_human})             & 572k      & 366k       & 5454        & 210    & 118          & 3735  & 704   & 22    & 11    & 4         \\
        & Lab (Fig.~\ref{fig:vs_human})              & 842k      & 610k	      & 13234	    & 215    & 123          & 5332  & 2463  & 38    & 72    & 53        \\
        & Apartment (Fig.~\ref{fig:vs_human})        & 411k	     & 310k	      & 10307	    & 131    & 112          & 2967  & 1353  & 17    & 17    & 3         \\
        & Highrise (Fig.~\ref{fig:vs_human})         & 337k	     & 235k	      & 5984	    & 103    & 152          & 3494  & 1520  & 26    & 42    & 19        \\

        \bottomrule
    \end{tabular}
\end{table*}

\paragraph{\textbf{Metrics \& Configurations}}

To measure the simplicity of generated LOD models, we define the simplification rate $s$ as the ratio between the number of output triangles and the number of input triangles. We also consider two metrics (\emph{steps} and \emph{cuts}) to evaluate the work involved in generating a LOD model $m\in M$. The \emph{cuts} metric represents the number of BSP-node splits needed during the traversal of the corresponding BSP-Tree in order to generate model $m$. The \emph{steps} metric counts the number of coarser LOD models in $M$ before reaching $m$.

To evaluate the accuracy of LOD models, we compute the Lowpoly~\citep{low-poly} metrics, denoted as $\tau_n$ and $\tau_s$, which quantify the visual difference between two meshes. Additionally, we measure the geometry fidelity in 3D using the RMSE distance, represented as a percentage of the diagonal length of the bounding box. The RMSE distance is computed bidirectionally, with $e_1$ representing the distance from the output LOD model to the input data and $e_2$ representing the distance from the input data to the output model.

\begin{figure*}[t!]
	\centering
	\includegraphics[width=1\linewidth]{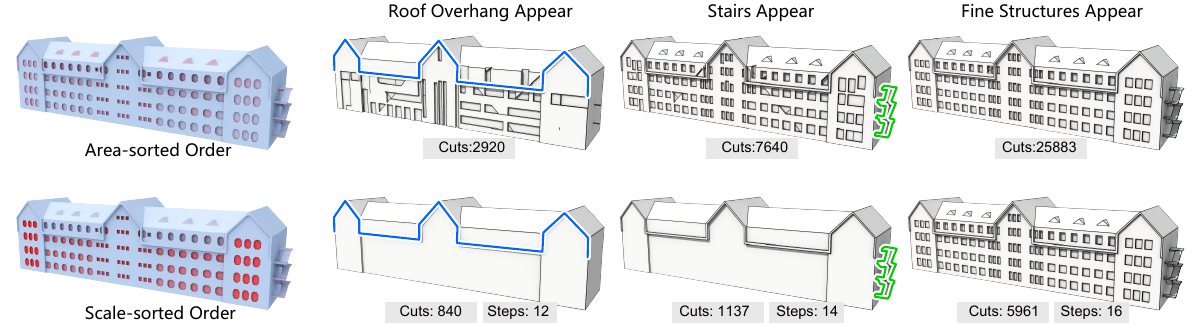}
	\caption{
		LOD-Tree vs BSP-Tree. 
		BSP-Tree selects planes to split based on their areas, resulting in small but visually/semantically important planes having low priorities. For example, the roof overhang (blue) and the exterior stairs (green) are modelled by tiny planes, which are considered low priority in BSP-Tree. Based on IO-View Analysis, LOD-Tree recognizes the importance of these planes and uses 840 cuts to model the roof overhang and 1137 cuts to model the stairs.
	}
	\label{fig:vs_BSP_Tree}
\end{figure*}

\begin{figure*}[t!]
	\centering
	\includegraphics[width=0.95\linewidth]{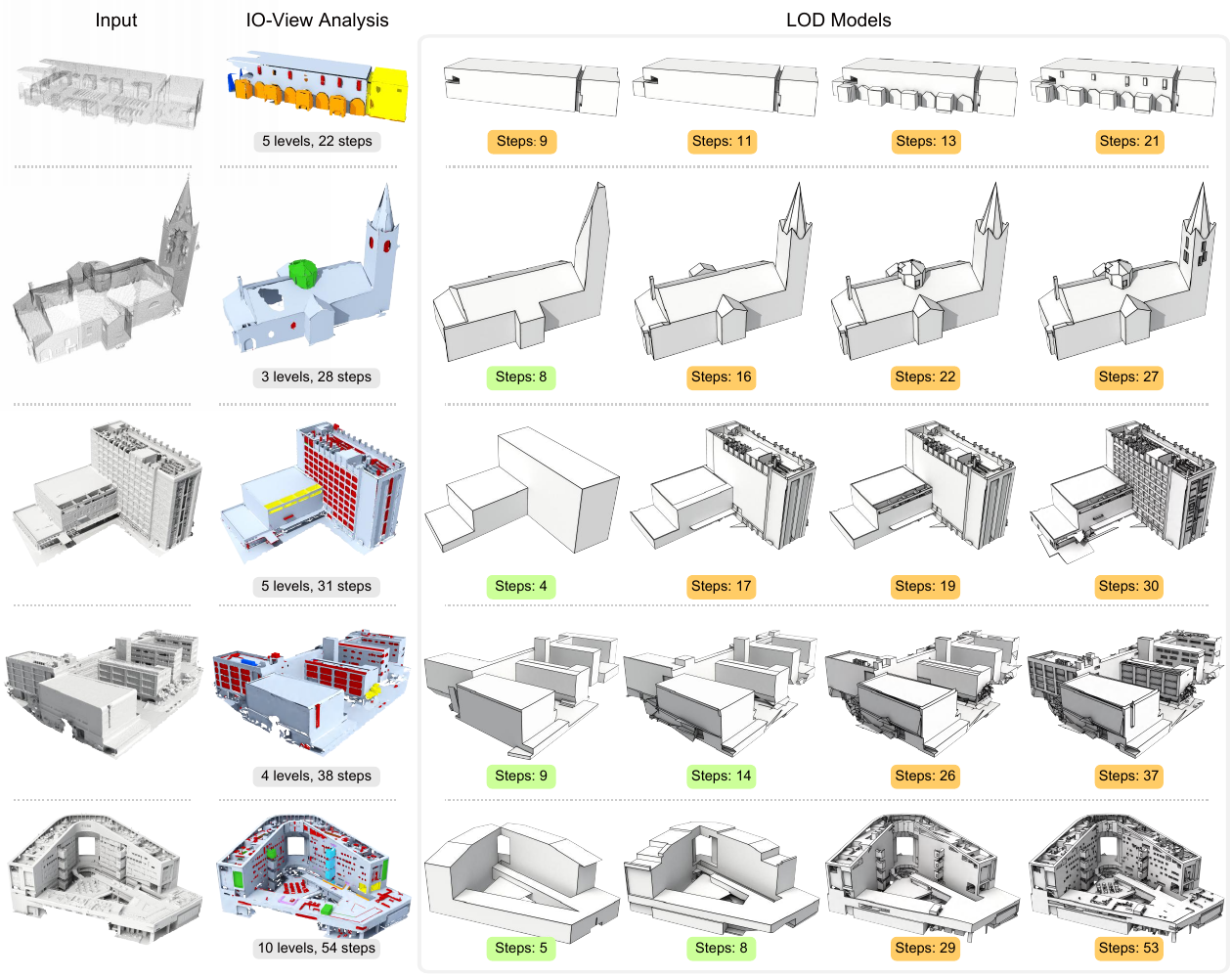}
	\caption{
		Results of real-world data. Our algorithm generates polygonal meshes with varying levels of detail for various datasets, including Church, Lans, Hall, School and CSSE. The principal planar primitives identified by IO-View are highlighted in blue, while the secondary planar primitives from different scale levels are displayed in different colors. Four LOD models are selected from the generated candidate LOD set $M$, where the models marked in orange are anchor models and those marked in green are interpolation models.
	}
	\label{fig:gallary}
\end{figure*}

Our algorithm is implemented in C++ using the CGAL library.
All experimental results presented are obtained on a workstation equipped with 
an Intel(R) Xeon(R) Gold 6100 CPU with 2.30 GHz and 191 GB RAM.

\subsection{LOD-Tree vs BSP-Tree}
\label{sec:ablation}

LOD-Tree has two improvements over traditional BSP-Tree: (1) better primitives order for space partition; (2) multi-branch nodes instead of double-branch nodes. These two improvements make our LOD-Tree more compact and structure-aware, as illustrated in Fig.~\ref{fig:vs_BSP_Tree}.

Traditional BSP-Tree-based methods often rely on the assumption that larger planar primitives have a broader impact on the final model. As a result, they sort the planar primitives based on area and perform sequential space partitioning. However, we observed that relying solely on the area of a planar primitive does not necessarily correlate with its visual and semantic significance. This is evident in the results of standard BSP-Tree space partitioning, \eg, Fig.~\ref{fig:vs_BSP_Tree}, where small planar primitives highlighted in blue do not appear until after approximately 2,920 cuts. However, these primitives, which belong to key architectural elements like roof overhangs, should logically emerge much earlier in the partition process. In contrast, our IO-View analysis identifies these planar primitives as principal primitives, resulting in their early appearance in the partition process (approximately 840 cuts compared to 2,920 cuts). The same observation can be made for the incoherent emergence of stairs (marked in green). Based on the IO-View, the LOD-Tree ensures the effective order of structure emergence and generates a much more meaningful model compared to the BSP-Tree at any intermediate state.

Beyond improving the emergence order of critical structures, the LOD-Tree exhibits a compact structure by automatically skipping many meaningless cuts based on the merging of BSP nodes, avoiding the generation of incomplete, broken structures. Fig.~\ref{fig:vs_BSP_Tree} demonstrates that the LOD-Tree only requires 5961 cuts to generate the final mesh, whereas the BSP-Tree involves 25,883 cuts to complete the space partition. Additionally, the simple traversal strategy provides a more efficient way to explore the LOD modelling space, \eg, it only takes 16 steps to generate all candidate LOD models.

\begin{figure*}[t!]
	\centering
	\includegraphics[width=1\linewidth]{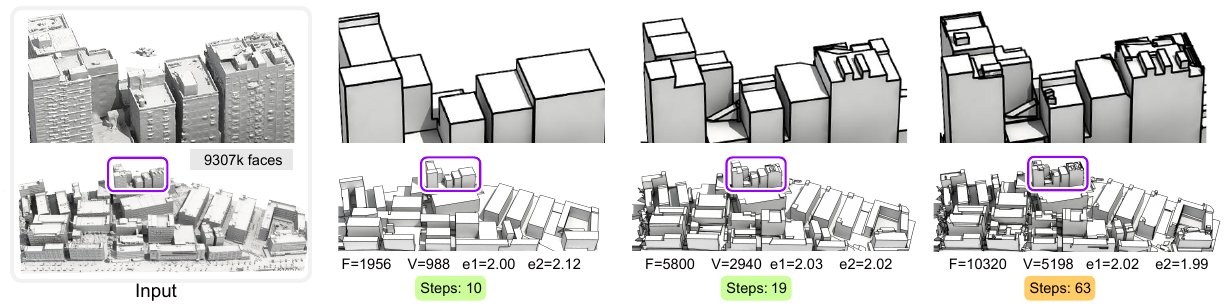}
	\caption{
		Experiment on a real complex scene: Residency. 
		Due to high building density and large-scale variation (sizes of windows vs. the whole scene), facade details are challenging to detect. 
		Nevertheless, roof details are nicely captured at proper LOD models. Hyper Parameters: $\epsilon$ = 0.5m, $\theta$ = 60$^\circ$, $\sigma$ = 500.
	}
	\label{fig:residence}
\end{figure*}

\subsection{Robustness to imperfect data}
\label{sec:robustness}

\paragraph{\textbf{Results of real-world data}}
The majority of input models in our dataset are reconstructed using UAV photogrammetry, which often introduces various artifacts such as non-manifold geometry, self-intersections, inaccuracies, and noise. However, our proposed LOD-Tree approach demonstrates its effectiveness in handling these challenges across different scenes.

\begin{figure}[t!]
	\centering
	\includegraphics[width=0.9\linewidth]{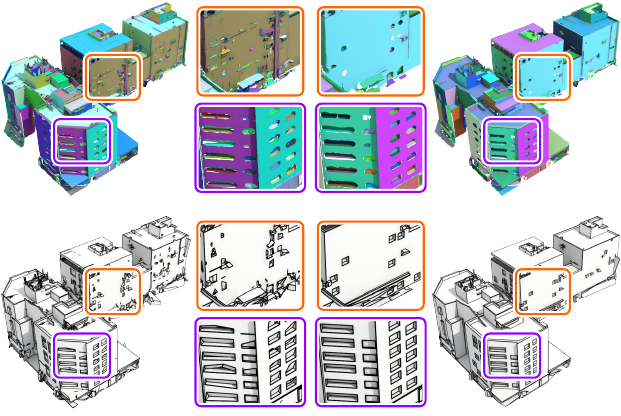}
	\caption{
		The impact of adaptive regularization.
		Plane detection inevitably generates redundant primitives due to noise, but most of these primitives can be eliminated through our adaptive regularization (highlighted in orange).
		Also, the accuracy of windows has been further improved by our template extraction (highlighted in purple).
	}
	\label{fig:Reg}
\end{figure}

In our urban reconstruction pipeline, plane detection plays a crucial role in preserving fine details. We employ small distance thresholds during plane detection and utilize adaptive regularization techniques to refine the results. Furthermore, our space partition process, based on the detected planes, exhibits resilience to occlusions. This allows us to recover missing parts by expanding the primitive to compensate for occluded regions. Fig.~\ref{fig:gallary} showcases our results on various datasets, demonstrating the ability to generate clean and accurate LOD models. 
And Fig.~\ref{fig:Reg} shows the impact of our adaptive regularization.

\paragraph{\textbf{Scalability}}
To demonstrate the scalability of our algorithm, we showcase the generation of LODs for large urban scenes. The LOD progression of a large-scale Residency model across three levels is shown in Fig.~\ref{fig:residence}. Initially, the roof is depicted as flat, but as the LODs progress, height information is gradually incorporated. Ultimately, the prismatic structure on the top is accurately recovered.

\paragraph{\textbf{Robustness to noise}}

\begin{figure}[t!]
	\centering
	\includegraphics[width=\linewidth]{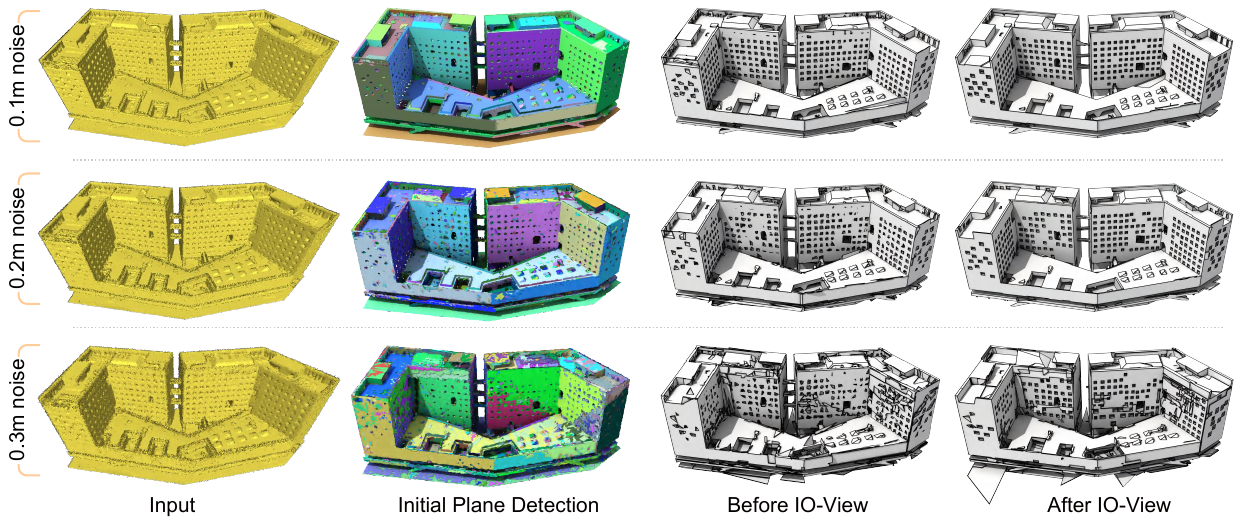}
	\caption{
		Robustness to noise. The impact of noise is hardly noticeable when planar primitives can be decently extracted from the input data (top 2 rows). However, when planar primitives cannot be reliably extracted, the quality of generated models also deteriorates.
		Compared to the finest models generated before IO-View, the models generated after IO-View are cleaner and have more regular details, such as windows and other secondary structures.
	}
	\label{fig:noise}
\end{figure}

\begin{figure*}[t!]
	\centering
	\includegraphics[width=1\linewidth]{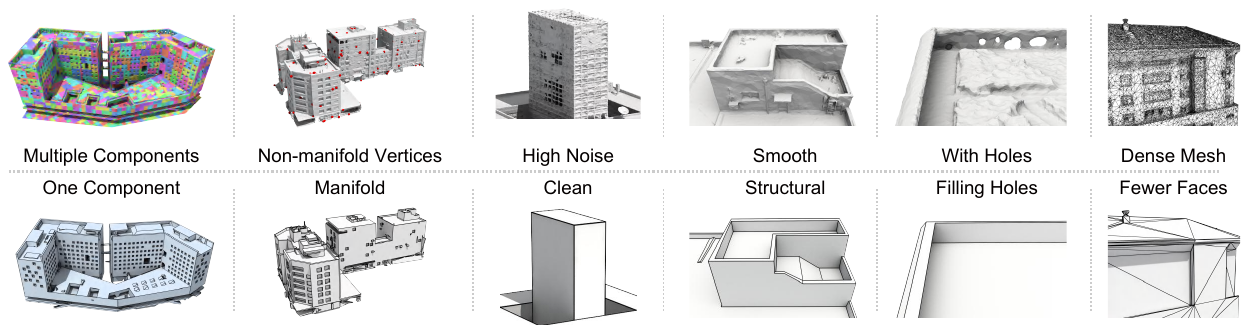}
	\caption{
		Comparisons between input and output meshes. The first row presents challenges commonly found in real-world data, such as multiple components, non-manifold vertices, high noise, smooth surfaces, holes, and dense, non-uniform triangle faces. In contrast, the second row demonstrates the outputs of our method, which successfully generates models with a single component, manifold vertices, clean geometry, and reduced face count. Our approach excels at preserving sharp features while filling holes and optimizing mesh complexity, demonstrating its robustness in handling diverse real-world scenarios. 
	}
	\label{fig:input}
\end{figure*}

Fig.~\ref{fig:noise} shows the robustness of our algorithm to noise. 
We introduced various levels of Gaussian noise deviation $\sigma$ (0.1m, 0.2m, 0.3m) to the point cloud.
When $\sigma \le 0.2m$, our algorithm outputs an accurate mesh. Plane detection is most easily affected by data quality. When $\sigma$ reaches 0.3m, some planar primitives will be lost or the detection will be inaccurate, resulting in less clean models, loss and deformation of secondary structures, and misclassification of secondary structures in the IO-View analysis. Since our algorithm can analyze and adaptively regularize these planar primitives, the impact of noise on plane detection and the final model is mitigated to a certain extent.
We also want to highlight that the input models are coming from real-world data, which contains all kinds of noise and artifacts that are far more complex than the synthetic noise we added. However, our algorithm can still generate clean and accurate LOD models for these real-world data, as shown in Fig.~\ref{fig:input}.

\paragraph{\textbf{User study}}

	To verify the usability of our tool for non-expert users, we conducted a user study comparing the generation of LOD models using Lowpoly~\citep{low-poly}, QEM~\citep{QEM} and our method. We used each of the three methods to generate candidate models at various LODs. This process was fully automated, requiring no user intervention.

	Since each method typically produces more candidate models than needed, users were asked to imagine the four most ideal LOD models and select them from each method's candidate set. We evaluated the methods based on their ability to generate the ideal LOD models envisioned by the users and the time required for users to select these models from the candidate set, especially when a method generates numerous candidates.
	
	A desirable method should produce a minimal number of candidate models while ensuring that the ideal LOD models are included in the candidate set. After selection, users rated the following statements on a scale from 0 to 10:
	\begin{itemize}
	\item Quality of models in level $i$: This method can generate good LOD models at level $i$.
	\item Precision: The method generates LOD models accurately, allowing easy selection of ideal models from the candidate LOD model set.
	\item Recall: Despite the number of candidate models, this method can generate all the models I envisioned.
	\item Overall performance: I would choose this method to generate LOD models.
	\end{itemize}
	Additionally, we recorded the average time users spent selecting the ideal LOD models from the candidate set for each input data. The results are shown in Table~\ref{table:user_study}. 
	Our method significantly outperforms Lowpoly and QEM across all metrics, achieving the highest scores in model quality (L1-L4), precision (P), recall (R), and overall performance (O), while also being the most time-efficient (T).

	\begin{table}[t!]
    \caption{
        The average scores (0-10) of different methods in a user study with 25 participants are reported below. Higher scores indicate better performance. We present the average scores for models at different levels, precision (P), recall (R), overall performance (O), and the time (T: min) required by the participant to pick the appropriate LODs for each input model.
    }
    \centering
    \label{table:user_study}
    \scalebox{0.8}{
    \begin{tabular}{cccccccccc}
        \toprule
         & Methods  & L1 & L2 & L3 & L4 & P  & R  & O & T  \\ \hline

         & Lowpoly & 5.73	& 5.66	& 5.52	& 5.77  & 5.02 & 5.33 & 5.02 & 2.81 \\
         & QEM      & 6.02	& 6.31	& 6.06	& 7.15  & 5.68 & 5.74 & 5.65 & 2.50
         \\
         & Ours     & \textbf{8.18}	& \textbf{8.20}	& \textbf{8.32}	& \textbf{8.45}  & \textbf{8.45} & \textbf{8.32} & \textbf{8.53} & \textbf{1.88} \\
        \bottomrule
    \end{tabular}
    }
\end{table}

\begin{figure}[t!]
	\centering
	\includegraphics[width=\linewidth]{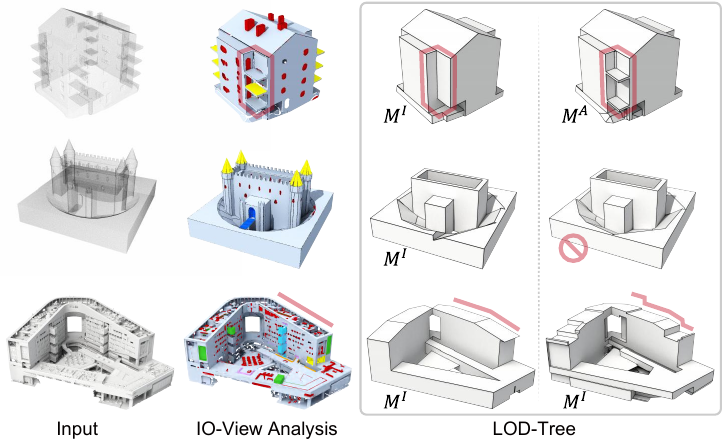}
	\caption{
		Failure cases of our method. The first row shows the situation where the secondary structure is embedded on multiple planes and is difficult to successfully extract by IO-View. The second row shows that LOD-Tree extracting the interpolation model through 80\% of the summed diff-value may lead to missing the optimal point (see the fourth column). The third row shows that due to the single configuration of plane detection, the stepped planes are difficult to further abstract into a simple inclined plane.
	}
	\label{fig:failure}
\end{figure}

\paragraph{\textbf{Failure cases}}
In some special cases where individual objects are connected via drainpipes or tubes or objects are situated at the junction of multiple planes, IO-View may encounter difficulties even with the $\alpha$-value adjusted. Fortunately, such challenges can be addressed through interpolation of the LOD-Tree. As shown in the first row of Fig.~\ref{fig:failure}, the balcony of the House Model is embedded on two facades, with parts identified as details (yellow) and parts as the principal structure (blue). However, even if the anchor model ($M^A$) is not ideal, we can still generate a LOD model ($M^I$) without balconies through interpolation. This interpolation mechanism allows us to handle complex cases and generate appropriate LOD representations.
On the other hand, our interpolation mechanism, which extracts models with a summed diff-value lower than 80\% of the previous one, sometimes misses the optimal point. As shown in the second row of Fig.~\ref{fig:failure}, LOD-Tree considers the model in the third column as an interpolation model, while skipping the better-looking model in the fourth column due to smaller changes from the previous model.
Additionally, for complex cases that do not contain embedding relations, such as the stepped structure shown in the third row of Fig.~\ref{fig:failure}, it is difficult for us to abstract it into a simple slope through interpolation. Since LOD-Tree is only based on plane detection with a single configuration, the parameters of the plane cannot be switched flexibly. In the future, we will consider how to integrate multiple configurations of plane detection into the construction of our LOD-Tree.

\begin{figure}[t!]
	\centering
	\includegraphics[width=1\linewidth]{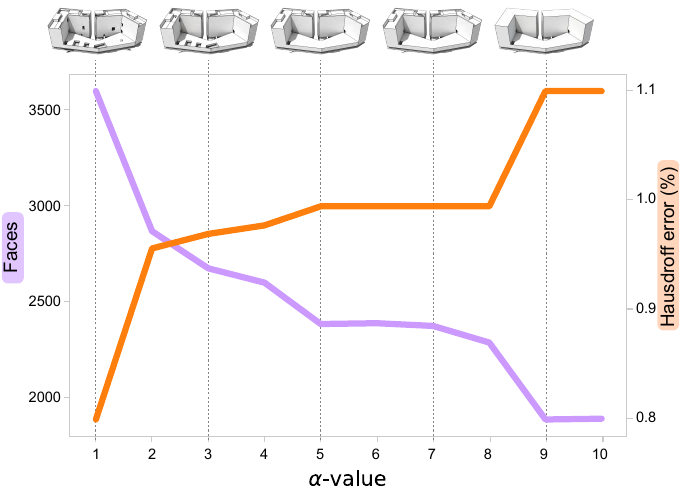}
	\caption{
		The impact of $\alpha$-value. Increasing $\alpha$ produces a more compact principal structure (purple curve), which allows better separation between principal and secondary structures. However, this increases the geometric error $e_2$ between the principal structure and the input (orange curve). Considering that building models usually have similar scales, setting the $\alpha$-value to 7m can effectively adapt to most cases.
	}
	\label{fig:alpha}
\end{figure}

\paragraph{\textbf{Ablation study}}
In this section, we conduct an ablation study to evaluate the influence of the $\alpha$-value on the extraction of the principal model, as well as the effect of the percentage threshold on the extraction of interpolation models. Additionally, we examine the role of the merging parameter $K$ in constructing the LOD-Tree to balance simplicity and expressiveness.

The $\alpha$-value affects the extraction of the principal and secondary structures; see Fig.~\ref{fig:alpha}. Through extensive experimentation, we have found that setting an $\alpha$-value of 7m can effectively adapt to most cases, considering that building models often exhibit similar scales. This parameter has been consistently utilized across all our experiments.

\begin{figure}[t!]
	\centering
	\includegraphics[width=\linewidth]{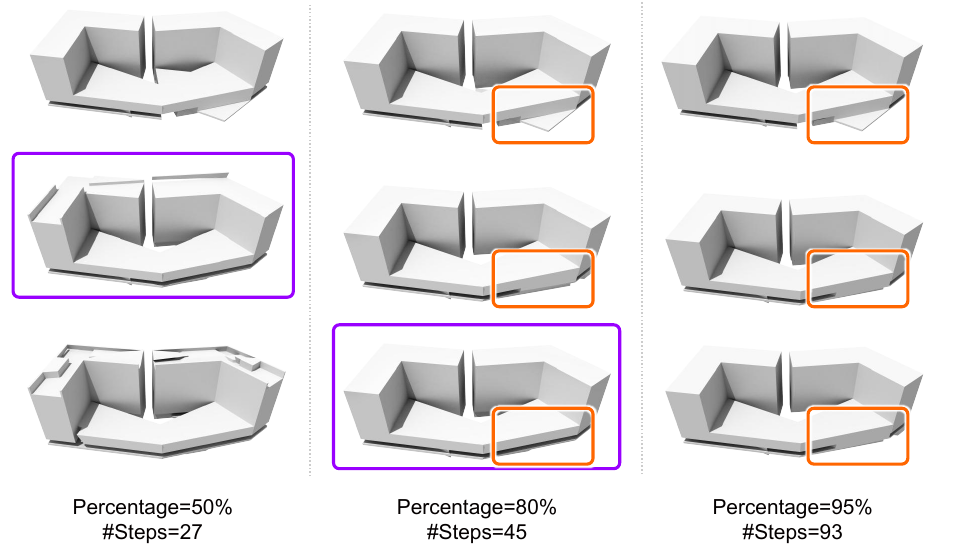}
	\caption{
		Impact of the percentage threshold on interpolation model extraction. A Higher percentage results in more interpolation models with smaller geometric differences between adjacent models (highlighted in orange), while a lower percentage produces fewer models but may skip optimal states (highlighted in blue). A percentage of 80\% strikes a balance between preserving optimal states and maintaining manageable interpolation complexity.
	}
	\label{fig:criteria}
\end{figure}

Next, we analyze the effect of varying the percentage on the extraction of the interpolation models as shown in Fig.~\ref{fig:criteria}. When the percentage is higher, more interpolation models are extracted, and the geometric differences between adjacent interpolation models become smaller. Conversely, a lower percentage results in fewer interpolation models being extracted, increasing the likelihood of skipping the optimal interpolation models. A percentage of 80\% achieves a good balance between facilitating user interaction and retaining the optimal interpolation models as much as possible. However, as noted in Fig.~\ref{fig:failure}, there is still a possibility that some key states may be skipped.

\begin{figure}[t!]
	\centering
	\includegraphics[width=\linewidth]{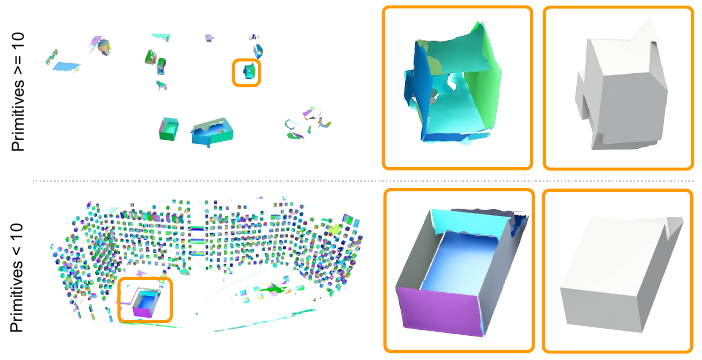}
	\caption{
		Impact of the merging parameter $K$ on the LOD-Tree construction. Structures with a larger number of primitives (e.g., $K$ $\geq$ 10) exhibit higher potential for further abstraction using the LOD-Tree traversal strategy, enabling more compact and concise representations. The BSP nodes corresponding to structures that are simple enough (e.g., $K$ < 10) and do not need to be interpolated are merged into one LOD node to generate a simpler LOD-Tree.
	}
	\label{fig:merging}
\end{figure}

Furthermore, we illustrate the impact of merging parameter $K$ on the LOD-Tree construction. We show secondary structures with primitive numbers $K$ < 10 and $K$ $\geq$ 10 in Fig.~\ref{fig:merging}, respectively. Structures with more primitives have greater potential for further abstraction. By traversing the LOD-Tree, these structures can be further interpolated, resulting in more abstract models. For a cluster containing fewer than 10 primitives per structure, we merge BSP nodes that correspond to the cluster's primitives into a single LOD-node. We set $K$ to 10 to balance the simplicity and expressiveness of the LOD-Tree.

\begin{figure*}[t!]
	\centering
	\includegraphics[width=1\linewidth]{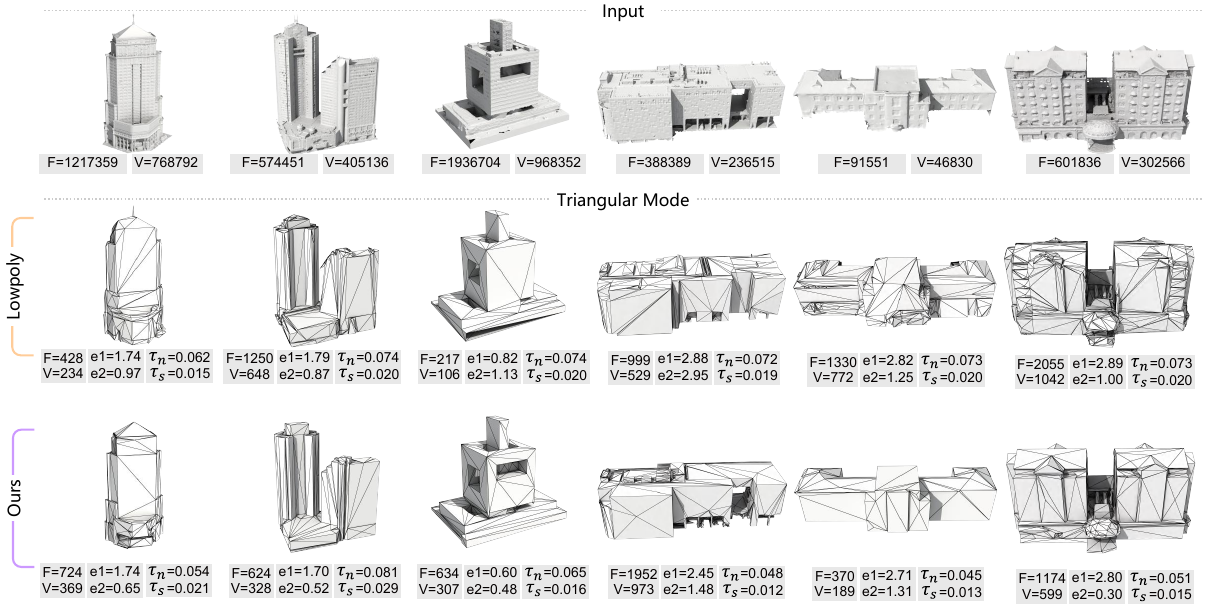}
	\caption{
		Comparisons with Lowpoly on six models (Headquarter, Department, Hitech, Mall, Office, Hotel). Both can generate concise models, but our method outperforms Lowpoly in terms of visual errors ($\tau_n$, $\tau_s$) and geometric errors ($e_1$, $e_2$). In particular, we show superior ability in recovering hollow structures in buildings (\eg, Hitech and Mall).
	}
	\label{fig:vs_low_poly}
\end{figure*}

\begin{figure}[t!]
	\centering
	\includegraphics[width=\linewidth]{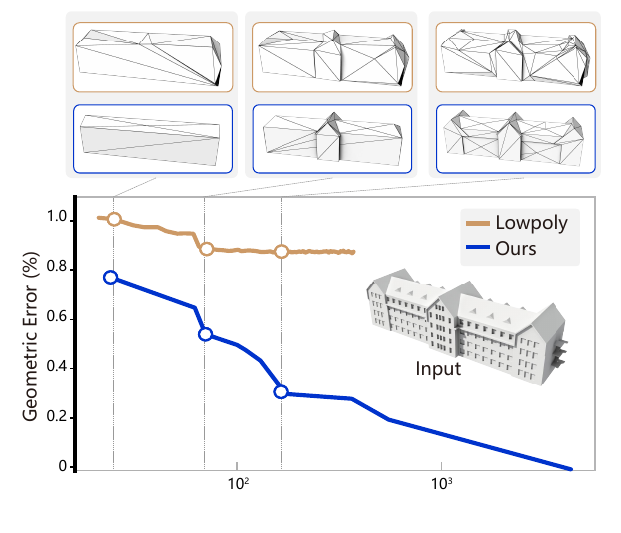}
	\caption{
		Comparison of the Pareto front sets between the Lowpoly and ours. Lowpoly focuses on extracting a simple mesh composed of a small number of faces. The mesh we extract spans the range of coarse models and fine models. Secondly, with the same number of faces, the geometric error of the mesh we extract is smaller than that of Lowpoly.
	}
	\label{fig:pareto}
	\vspace{-10pt}
\end{figure}

\subsection{Comparisons with visual-dependent LOD methods}
\label{sec:sota_visual}

In this section, we compare the performance of our LOD models with alternative LOD algorithms that focus on visual-dependent criteria. We consider three different pipelines:
(1) Method based on the view contour, like Lowpoly~\citep{low-poly}, which is guided by two-dimensional pixel errors; 
(2) Learning-based method, like NeuralLOD~\citep{neuralLOD}, which is guided by the resolution of implicit field; 
and (3) Classic edge collapse algorithm, such as QEM~\citep{QEM} and Robust-lowpoly~\citep{robust-lowpoly}, which is guided by geometric errors.
To evaluate the performance, we recorded the average processing time required for all input models across the methods, as shown in Table~\ref{table:time}. 

\begin{table}[t!]
    \caption{
        Average time compared with visual-dependent LOD methods. 
    }
    \centering
    \label{table:time}
    \scalebox{0.85}{
        \begin{tabular}{cccccccccccccc}
            \toprule
            &              & Lowpoly  & NeuralLOD  & QEM        & Robust-lowpoly  & Ours  \\ \hline 
            
            & $T$          & $351 s$      & $>1 h$     & $<20 s$    & $402 s$         & $294 s$   \\
            \bottomrule
        \end{tabular}
    }
\end{table}

\begin{figure*}[t!]
	\centering
	\includegraphics[width=1\linewidth]{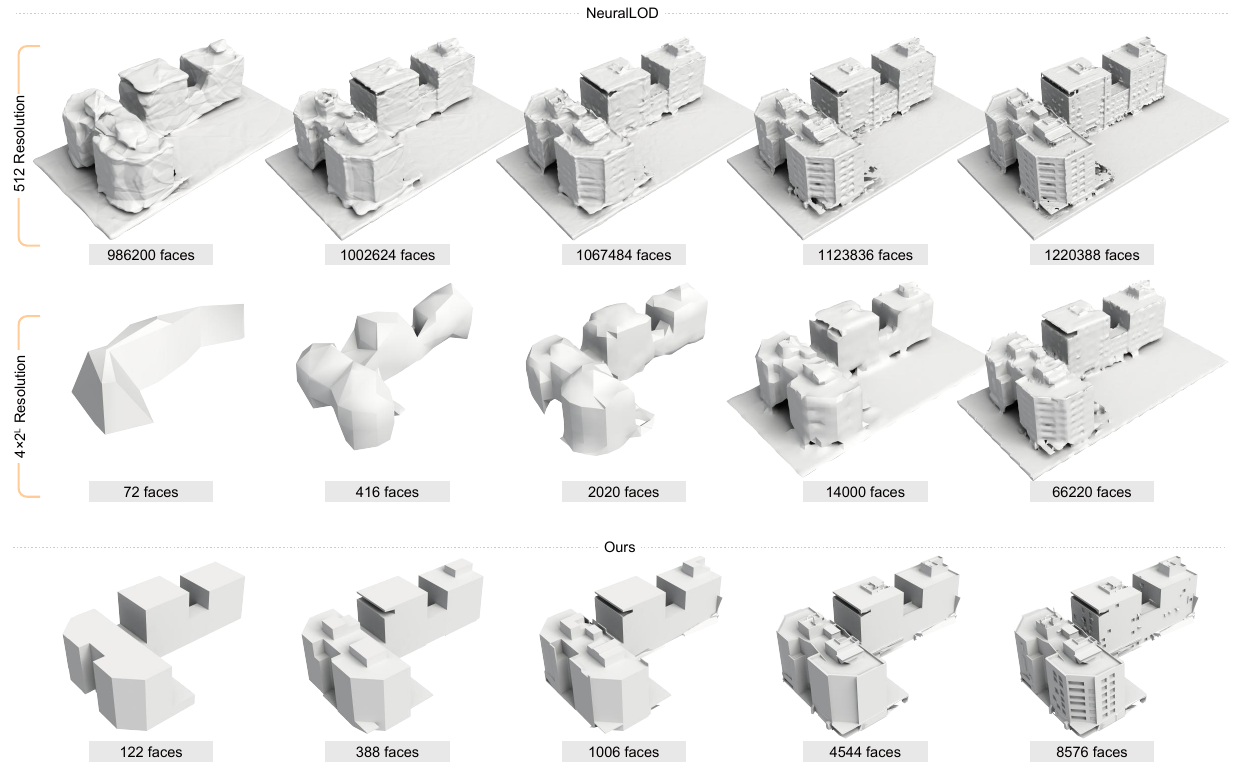}
	\caption{
		Comparisons with NeuralLOD~\citep{neuralLOD} on AviTech model.
		NeuralLOD tends to smooth geometric details between LODs and has no hard constraints on the preservation of sharp features. We show the results of reconstructing the implicit field of NeuralLOD with Marching Cubes at 512-resolution and $4*2^L$-resolution respectively. NeuralLOD cannot strike a good balance between fidelity and simplicity. In contrast, our results capture almost all visually important structures, such as the sinuous principal structure and most of the details, while having fewer faces.
	}
	\label{fig:vs_NeuralLOD}
\end{figure*}

\begin{figure*}[t!]
	\centering
	\includegraphics[width=1\linewidth]{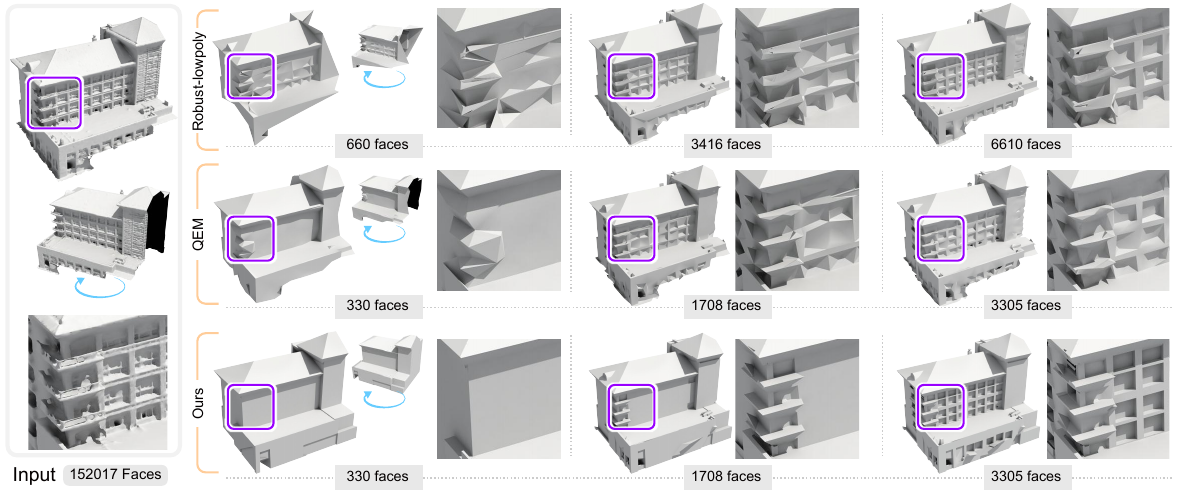}
	\caption{
		Comparison with QEM and Robust-lowpoly. While being able to generate models with increasing LOD, QEM struggles to distinguish the principal and secondary structures of the model and is unable to complete missing parts, as shown in the results of another view indicated by the blue arrow. Robust-lowpoly can generate a watertight closed mesh and better maintain the sharp features of the data, but the boundaries will gradually shrink during the simplification process. In contrast, our algorithm exhibits robustness against geometric and topological errors in the input and can fill in missing parts, allowing us to produce models that are watertight, highly accurate, and structurally faithful.
	}
	\label{fig:QEM}
\end{figure*}

\paragraph{\textbf{Comparisons with Lowpoly}}
In order to compare with the Lowpoly method~\citep{low-poly}, we processed our data using Lowpoly's publicly released executable program. However, due to the slow processing speed of Lowpoly on large models, it was challenging to handle input models with hundreds of thousands of faces. Therefore, we used the most refined model generated by our algorithm as the input for Lowpoly and selected the model from their Pareto front set that maintains the smallest error but has as few faces as possible for comparison. As shown in Fig.~\ref{fig:vs_low_poly}, Lowpoly struggles to preserve the structure of the models. To ensure a fair comparison, we selected the interpolation models from our approach that closely matched the structural information of Lowpoly's output. Both can produce compact models, but our method outperforms Lowpoly in terms of visual errors ($\tau_n$, $\tau_s$) and geometric errors ($e_1$, $e_2$). In particular, we demonstrate superior capability in recovering hollow structures in buildings, as exemplified by the Hitech and Mall models in Fig.\ref{fig:vs_low_poly}.

Additionally, we analyze the differences between the Pareto front sets extracted by both methods as shown in Fig.~\ref{fig:pareto}. Our approach generates models spanning multiple levels of detail, from coarse to fine, while Lowpoly focuses primarily on compact representations. As a result, our models cover a wider range of face counts. Even at the same face count, our method achieves significantly lower geometric errors, highlighting its effectiveness and versatility in producing high-quality results.

\paragraph{\textbf{Comparisons with NeuralLOD}}
NeuralLOD controls the levels of detail through the depth of an octree, while our approach controls the levels through the depth of our LOD-Tree. To extract meshes for comparison, we reconstruct the implicit field of NeuralLOD with Marching Cubes~\citep{lorensen1987marching} at 512-resolution and $4*2^L$-resolution respectively.
As shown in Fig.~\ref{fig:vs_NeuralLOD}, the transition between the models generated by NeuralLOD is smoother, but it is difficult to maintain planarity and sharp features. It is more suitable for free-form models. For more structural architectural models, the LODs generated by our LOD-Tree are more concise and more structure-aware.

\begin{figure*}[t!]
	\centering
	\includegraphics[width=1\linewidth]{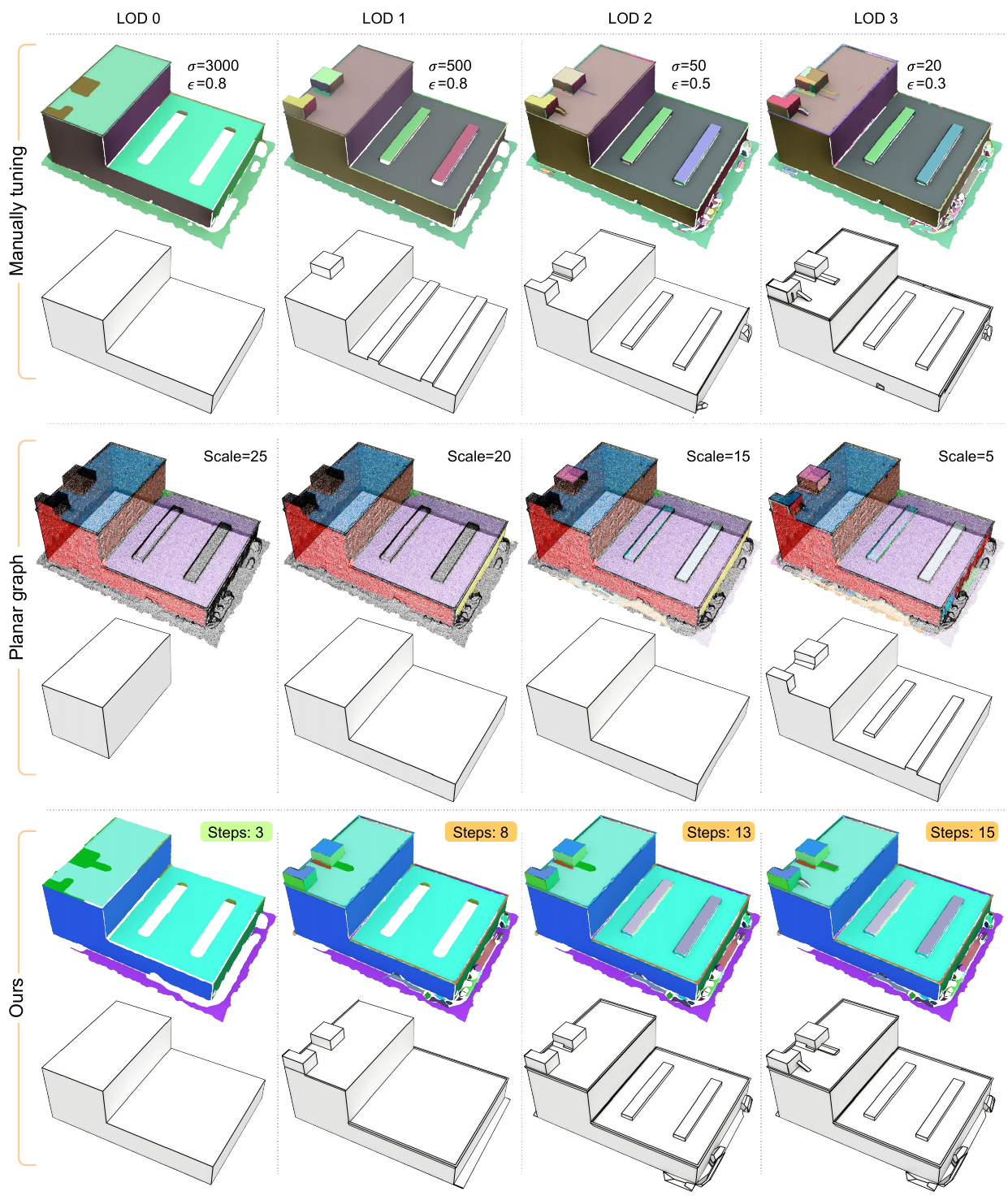}
	\caption{
		Comparisons of LOD models generated by different multi-scale plane detection methods. 
		Our IO-View analysis yields the best result in terms of accuracy and flexibility.
		Planar graph~\citep{LejemblePersistence} and manually tuning parameters failed to extract the correlation between planar primitives, thus generating meaningless and inaccurate details.
	}
	\label{fig:vs_structural}
\end{figure*}

\begin{figure*}[t!]
	\centering
	\includegraphics[width=1\linewidth]{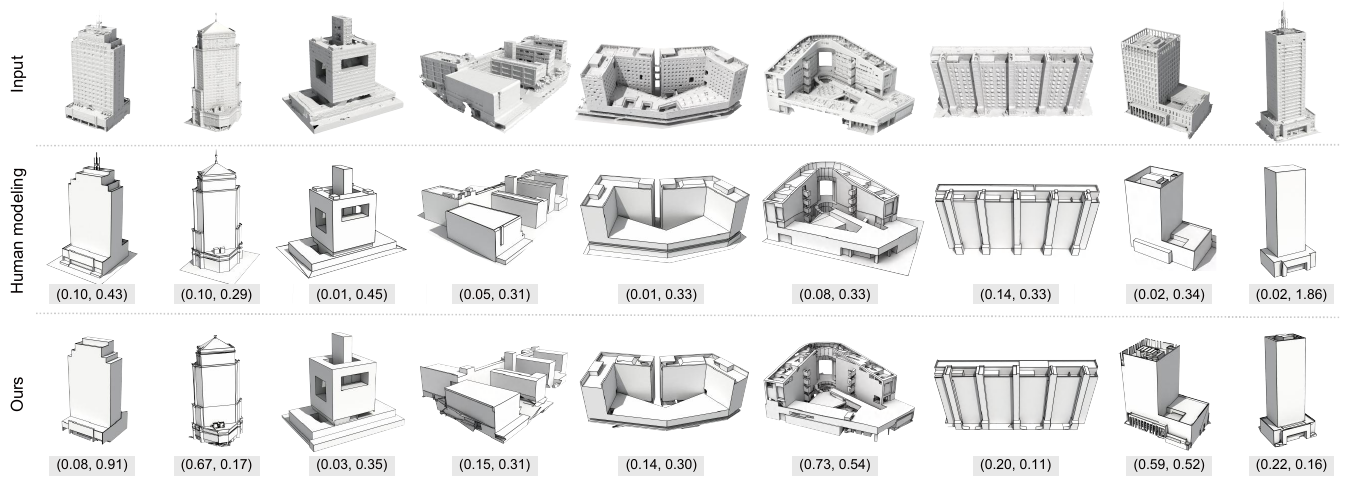}
	\caption{
		Comparisons with human-created models. 
		Since plane detection does not handle noise as well as humans, our simplification rate is higher than humans even for visually similar models.
		However, due to the fidelity of plane detection to the data, the error of our model can be similar to or even better than that of human-created models.
		(•,•) denotes (simplification rate $s$, RMSE $e_1$).
	}
	\label{fig:vs_human}
\end{figure*}

\paragraph{\textbf{Comparisons with QEM and Robust-lowpoly}}
Furthermore, we compare with QEM and the state-of-the-art Robust-lowpoly. Robust-lowpoly is capable of generating a watertight closed mesh through remeshing. However, enforcing watertight models may double the number of faces needed under the worst-case scenarios. 
Therefore, for a fair comparison, we allow their models to use twice as many faces. As depicted in Fig.~\ref{fig:QEM}, the QEM method encounters difficulties in preserving the boundaries and sharp features, while Robust-lowpoly can handle more challenging data and better maintain the sharp features due to the remeshing. However, Robust-lowpoly also struggles to maintain boundaries since it does not impose constraints on the boundaries during the simplification. In contrast, our algorithm demonstrates resilience in geometric and topological errors present in the input and can compensate for missing parts, enabling us to generate watertight, more accurate, and structurally faithful models.

\subsection{Comparisons with multi-scale plane detection}
\label{sec:sota}

    Existing structural reconstruction cannot directly produce LOD models. 
    However, we can mimic the LOD generation by generating several sets of planes at different scales combined with structural reconstruction.
    Intuitively, the coarser plane might produce the model in coarser levels.
    We generate multi-level planes by three methods:
	(1) manually tuning a set of parameters to obtain several sets of planes at different scales; 
	(2) constructing multi-scaled planar graph~\citep{LejemblePersistence}; 
	and (3) analyzing planes through our IO-View analysis.
    These planes are then fed into structural reconstruction methods KSR~\citep{KSR} to reconstruct models at different scales.

    Specifically, as for the first method, we manually tune the parameters $\sigma$ and $\epsilon$ provided by KSR to control the scale of the detected planes. $\sigma$ controls the minimum points required to generate a valid plane primitive, and $\epsilon$ controls the minimum distance required to classify whether points are inliers. 
	And for the planar graph~\citep{LejemblePersistence}, we follow the setting in their paper and extract planes in scale 5,15,20,25. 
	Finally, we set our plane detection parameters to be consistent with those of the finest level in the first method ($\sigma$=20, $\epsilon$=0.3m), and use IO-View to extract multi-level planes.

    Fig.~\ref{fig:vs_structural} shows models reconstructed by different methods. Our method generates the best results in terms of accuracy and flexibility. Manually adjusting parameters and the planar graph cannot efficiently extract the relation between different primitives, thus they can not select the planes properly to form meaningful and accurate LOD models.
	
\subsection{Comparisons with human modelling}
\label{sec:human}

Finally, we compare our reconstruction with human modelling, as shown in Fig.~\ref{fig:vs_human}. We ask experienced artists to create a single LOD model for each scanned point cloud or MVS mesh in our dataset, preserving what they consider the principal structures. We then pick the visually closest model from our generated set of LOD models to compare in terms of generation time, simplification rate, and RMSE ($e_1$).
Human-created models of varying complexity take from 1 hour to 1 day, while we only require a maximum of 30 minutes from generation to manual selection of LOD models. 
Plane detection can alleviate noise, but it still cannot remove all noise as correctly as humans.
Therefore, when the quality of input is poor, humans are able to approximate noisy surfaces with single planes whereas the plane detection algorithm often uses multiple planes. As a result, our simplification rate is higher than the results of human-created models, even though visually the models are very similar.
On the other hand, the automatic plane detection algorithm can approximate surfaces with greater accuracy than humans, resulting in lower RMSE errors for our results.
This indicates that our pipeline is ready to generate LOD models of sufficient quality and is also conducive for artists to fine-tune our generated models or add rich secondary structures on top of a lower complexity model.

\section{Conclusion and Future Work}

    We propose a new representation, LOD-Tree, for extracting models with different LODs from the dense point cloud or mesh.
    The LOD-Tree balances high-level semantic-based and low-level geometric-cues LOD methods while preserving the ability to manipulate the generated model efficiently. The success of the LOD-Tree relies on the holistic analysis of the planar primitives detected from input by the proposed IO-View, which reduces semantically meaningless states that may occur during the space partition process.

    \paragraph{Limitation} The core objective of our work is to address the challenge of extracting structural elements from unlabeled building datasets using a geometry-driven approach. While our proposed method leverages the interplay between planes and space partitioning, its effectiveness is inherently limited by the accuracy of the detected planes, which are more sensitive to noises commonly introduced to input models by MVS methods. Additionally, once the LOD-Tree is constructed, no mechanisms for further editing or modification are incorporated.
    
    As a future direction, we aim to explore how planes can be simultaneously detected, edited and regularized during the space partitioning process to develop a more comprehensive LOD representation. Moreover, our input data, derived from MVS point clouds generated from drone-captured images, is susceptible to noise. Investigating how to efficiently utilize image information directly for structure extraction could further reduce noise and enhance the accuracy of the generated LOD models.
    \vspace{2em}
    
    Funding: This work was supported in parts by National Key R\&D Program of China (\seqsplit{2024YFB3908500}, \seqsplit{2024YFB3908502}),  ICFCRT (\seqsplit{W2441020}), NSFC (\seqsplit{U21B2023}), Guangdong Basic and Applied Basic Research Foundation (\seqsplit{2023B1515120026}), Shenzhen Science and Technology Program (\seqsplit{KJZD20240903100022028}, \seqsplit{KQTD20210811090044003}, \seqsplit{RCJC20200714114435012}), and Scientific Development Funds from Shenzhen University.

\appendix
\section{Definitions of Key Terminologies}

In this section, we provide detailed definitions of the key terms and concepts used throughout this paper. This centralized glossary aims to facilitate the understanding of the relationships among these terms.

\paragraph*{\textbf{Input model $I$}}
The input data for our method consists of either oriented triangle meshes or point clouds with per-point normals, primarily acquired through multi-view stereo reconstruction from images captured by a single-camera drone.

\paragraph*{\textbf{Planar Primitives $P = \{p_i\}_{i=0}^N$}}
Planar primitives are planar regions detected through region growing techniques. Each planar primitive consists of a collection of points that are approximately coplanar. After the execution of the IO-View algorithm, planar primitives are further categorized into principal primitives and secondary primitives.

\paragraph*{\textbf{$\alpha$-shapes}}
$\alpha$-shapes are the bounded plane of each planar primitive, where the shape's boundary is controlled by the chosen $\alpha$-value. A smaller $\alpha$-value captures finer details and concavities, while a larger $\alpha$-value results in more convex boundaries.

\paragraph*{\textbf{Polyhedra / Leaf Nodes}}
A polyhedron, also referred to as a leaf node, is the final result of a binary space partitioning (BSP) process. It represents the smallest indivisible spatial unit within the BSP tree.

\paragraph*{\textbf{Regions}}
A region is determined by the spatial connectivity of polyhedra. Two adjacent polyhedra can be merged if and only if two conditions are met: first, they have the same in/out label, and second, they are not separated by any $\alpha$-shape. Connected polyhedra collectively form a single region. Regions are classified into four categories: Core interior $V_{in}$, Core exterior $V_{out}$, Addon structures $\{V_{+}\}$ and Cutout structures $\{V_{-}\}$.

\paragraph*{\textbf{3D Structures}}
A 3D structure refers to a solid block with a specific meaning of a building. These structures are categorized into principal structures and secondary structures.

\paragraph*{\textbf{Clusters $C=\{C_i\}_{i=0}^{N_c}$}}
Clusters are groups of secondary structures that are aggregated based on semantic similarity. All secondary structures are grouped into $C$ clusters, where each cluster $C_i$ represents structures with similar semantic meanings, such as doors or windows.

\paragraph*{\textbf{Level Sets $L=\{L_i\}_{i=1}^{N_l}$}}
Level sets are higher-level groupings of clusters. Clusters within the same level set share similar importance concerning visual appearance or practical applications. For example, in CityGML, doors and windows might be classified into the same level set.

\paragraph*{\textbf{Scale-sorted planar primitives $S=\{S_i\}_{i=0}^{N_l}$}}
Scale-sorted planar primitives refer to a set of primitives ordered by their structural significance. The principal primitives, denoted as $S_0$, are given the highest priority, followed by the secondary primitives $\{S_{1}, ..., S_{N_l}\}$, each corresponding to the primitives contained in levels $\{L_{1}, ..., L_{N_l}\}$, respectively.

\paragraph*{\textbf{Candidate LOD Model Set $M$}}
The candidate LOD model set $M$ consists of anchor models $M^A$ and interpolation models $M^I$. Anchor models are extracted once each level set completes space partitioning, while interpolation models are extracted during the space partitioning to capture significant changes in appearance.

\paragraph*{\textbf{Relationships Among Key Concepts}}
The relationships among principal primitives, secondary primitives, principal structures and secondary structures, are foundational to our model:

\begin{figure*}[t!]
	\centering
	\includegraphics[width=1\linewidth]{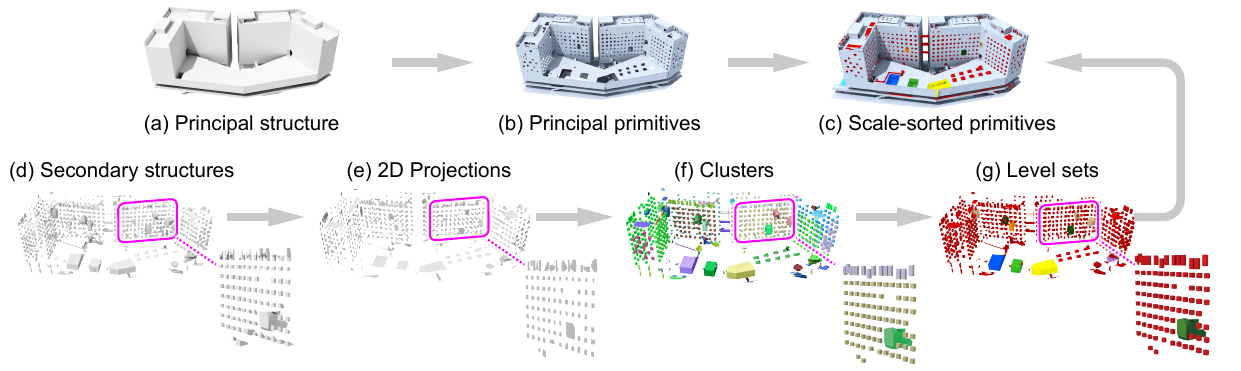}
	\caption{
		Illustration of key terminologies of IO-View Analysis. We begin by extracting the principal structure (a) and secondary structures (d) from the input geometry. A first-stage clustering is then performed based on the projected area (e) of secondary structures, resulting in a set of clusters (f), which are further regularized using our specific regularization strategy. Subsequently, a second-stage clustering groups the clusters based on volume size, forming level sets (g). Finally, principal primitives (b) and primitives of level sets are arranged in order to form the scale-sorted primitives (c), which are used for LOD generation.
	}
	\label{fig:terminologies}
\end{figure*}

\begin{itemize}
	\item The $\alpha$-shapes that separate regions $V_{in}$ and $V_{out}$ constitute the principal primitives.
	\item These principal primitives enclose the principal structure.
	\item The regions $\{V_{+}\}$ and $\{V_{-}\}$ form the secondary structures.
	\item Secondary primitives enclose the secondary structures.
\end{itemize}

Furthermore, the clustering of secondary structures influences the construction of the LOD-Tree as follows:
\begin{itemize}
	\item Since there is a correspondence between structures and primitives, clustering and sorting of structures inherently involve the clustering and sorting of primitives.
	\item The clustering and sorting of primitives determine the sequence of space partitioning of the LOD-Tree and indicate which BSP nodes are combined to form an LOD-node.
	\item This hierarchical organization ensures that the LOD-Tree effectively represents the geometric and semantic complexity of the building model, facilitating various downstream applications.
\end{itemize}

In Fig.~\ref{fig:terminologies}, we further provide illustrative legends to explain the meaning of each key terminology.

\section{Binary space partitioning}
\label{sec:bsp}
Given an initial set of planar primitives $P = \{p_i\}_{i=0}^N$ extracted by region growing~\citep{RegionGrowing} (default detection parameters: $\epsilon$ = 0.15m, $\theta$ = 40$^\circ$, $\sigma$ = 15), we use these primitives to partition the space and generate the BSP-Tree. 

As a preliminary step, we compute the convex hull of each planar
primitive, i.e. the smallest convex polygon containing the projection of
the inliers (points or triangles) on the optimal plane of the primitive.

The binary space partitioning begins with the recursive division of the bounding box of the model. Each primitive partitions a space into two subspaces. Each time we select the convex hull with the largest area, partition the subspace into two. During the operation, unprocessed convex hulls are assigned to their corresponding subspace. If a convex hull spans both subspaces, it is split into two to ensure that each convex hull in the new subspace remains within its designated region. This partitioning process continues until there are no more subspaces that can be divided. At the end of this process, each leaf node in the BSP-Tree corresponds to a convex polyhedron cell, and all the leaf nodes are combined to form the initial bounding box of the model. We use a 3D combinatorial map ~\citep{damiand2014combinatorial} to represent the partitioning results. The 3-cells of the combinatorial map correspond to the polyhedra (leaf nodes). And the 2-cells of the combinatorial map correspond to the polyhedra faces. For the region growing and combinatorial map, we use the implementation from the CGAL library\footnote{\url{https://doc.cgal.org/latest/Shape_detection/index.html}}.

\section{Polyhedra in/out labelling}
\label{sec:labelling}

\begin{figure}[t!]
	\centering
	\includegraphics[width=0.75\linewidth]{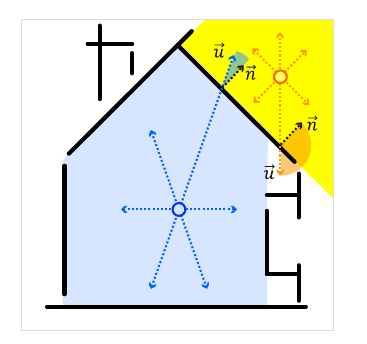}
	\caption{
		Illustration of the in/out labelling. The black line segments represent the detected $\alpha$-shapes. Each polyhedron emits 100 rays from the center. If $\vec{u}*\vec{n}>0$, the ray tends to mark the polyhedron as $in$, otherwise, it tends to mark it as $out$. Consequently, the blue polyhedron is labelled as $in$ and the yellow polyhedron is labelled as $out$.
	}
	\label{fig:raycasting}
\end{figure}

Once BSP is finalized, we perform polyhedra in/out labelling based on a ray-casting strategy~\citep{raystabbing}, as shown in Fig.~\ref{fig:raycasting}. Specifically, we uniformly emit 100 rays from the center of each polyhedron. For each ray $\vec{u}$ emitted by each polyhedron, find the triangle in the $\alpha$-shapes that the ray intersects for the first time, and calculate the inner product of the ray $\vec{u}$ and the normal $\vec{n}$ of the triangle. If $\vec{u}*\vec{n}>0$, the direction of the ray is consistent with the direction of the triangle, and it tends to mark the polyhedron as $in$ (assuming that the normal of all triangles point to the outside of the input), otherwise it tends to mark it as $out$. In addition, if the ray does not intersect any triangle, it also tends to mark the polyhedron as $out$. If more than half of the rays tend to mark the polyhedron as $in$, it is set as $in$, otherwise, it is set as $out$.

\section{Quantitative evaluation}

\begin{table*}[t!]
    \newcommand{\tabincell}[2]{\begin{tabular}{@{}#1@{}}#2\end{tabular}}
    \caption{
        Quantitative evaluation of different LOD generation methods across all datasets. \textbf{Bold}: best result; \underline{Underline}: second best. Geometric errors are computed with respect to the original inputs, which may contain noise or limited precision. As a result, methods that preserve input geometry more closely may report lower errors. Our method, which incorporates structural awareness, may exhibit slightly higher errors due to abstraction from noisy details while aiming to retain salient structural features.
    }
    \centering
    \label{table:metrics}
    \scalebox{0.95}{
    \begin{tabular}{c|ccc|ccc|ccc|ccc}
        \toprule
        \multirow{2}{*}{Methods} & \multicolumn{3}{c|}{LOD 0} & \multicolumn{3}{c|}{LOD 1} & \multicolumn{3}{c|}{LOD 2} & \multicolumn{3}{c}{LOD 3} \\
        \cmidrule(lr){2-4} \cmidrule(lr){5-7} \cmidrule(lr){8-10} \cmidrule(lr){11-13}
        & $F$ & $e_1$ & $e_2$ & $F$ & $e_1$ & $e_2$ & $F$ & $e_1$ & $e_2$ & $F$ & $e_1$ & $e_2$ \\
        \midrule
        Lowpoly         & 202       & 0.79       & 1.19        & -            & -        & -        & -           & -         & -        & -           & -         & -         \\
        NeuralLOD (512)           & 1032k     & \textbf{0.56}       & \textbf{0.68}        & 1064k        & 0.57     & \underline{0.50}     & 1106k        & 0.68        & 0.42       & 1182k        & 0.72        & 0.34        \\
        NeuralLOD ($4*2^L$)       & 71        & 6.64       & 11.77       & 501          & 2.75     & 6.25     & 2758         & 1.62        & 3.14       & 13795        & 1.06        & 1.76        \\
        Robust-lowpoly  & 928       & 0.75       & 1.09        & 3712         & \textbf{0.26}     & \textbf{0.36}     & 9563        & \textbf{0.15}      & \textbf{0.17}     & 21415       & \textbf{0.13}      & \textbf{0.15}        \\
        \hline\hline
        Manually tuning & 279       & 0.71       & \underline{1.07}        & 952          & 0.50     & 0.55     & -           & -         & -        & -           & -         & -        \\
        Planar graph    & 40        & 2.80       & 4.03        & 130          & 1.55     & 1.92     & 518         & 0.94      & 1.23     & 2151        & 0.57      & 0.64      \\
        Ours            & 297       & \underline{0.62}       & 1.13        & 1646         & \underline{0.42}     & 0.65     & 4498        & \underline{0.39}      & \underline{0.39}     & 10383       & \underline{0.37}      & \underline{0.28}       \\
        \bottomrule
    \end{tabular}
    }
\end{table*}

\begin{figure}[]
	\centering
	\includegraphics[width=0.95\linewidth]{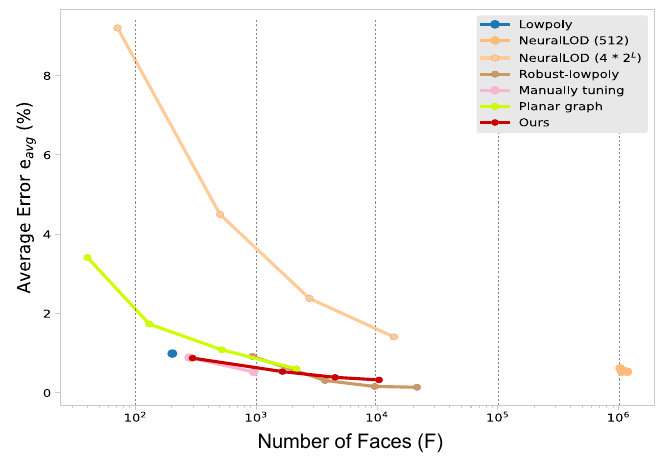}
	\caption{
		Illustration of average geometric error across different LOD generation methods. The average geometric error $e_{avg}$ is defined as the mean of $e_1$ and $e_2$ in Table~\ref{table:metrics}. Our method demonstrates overall competitive performance. Robust-lowpoly tends to preserve input geometry more closely at fine-level LODs due to its greedy, error-minimizing collapse strategy. In contrast, our global structure-aware approach maintains better geometric fidelity at lower face counts, highlighting its strength in coarse-level LOD generation.
	}
	\label{fig:errors}
\end{figure}

To comprehensively evaluate our method and highlight its differences from existing approaches, we conduct quantitative experiments across all datasets. Specifically, we select four representative levels of detail (LODs) and compare our method against four types of baselines: NeuralLOD, RobustLowpoly, and two multi-scale plane detection methods integrated with KSR reconstruction — a manually tuned method and a planar graph-based method. Since Lowpoly is specifically designed for producing models at LOD 0, we only include it in comparisons at that level. Additionally, due to the limitations of QEM in handling inputs with highly disconnected mesh topology, we report results only for the more robust variant, RobustLowpoly.

As shown in Table~\ref{table:metrics} and Fig.~\ref{fig:errors}, our method achieves competitive performance across all LODs. While the geometric errors ($e_1$, $e_2$) of our method are slightly higher in some cases, it is important to note that these metrics are computed with respect to input models that may contain noise or limited precision. Rather than strictly fitting such input, our method emphasizes structural coherence and semantic interpretability, which may result in modest deviations in error metrics but better capture the underlying geometry in a more meaningful manner, as shown in Fig.~\ref{fig:vs_low_poly}, Fig.~\ref{fig:vs_NeuralLOD}, Fig.~\ref{fig:QEM}, Fig.~\ref{fig:vs_structural}.

\bibliographystyle{elsarticle-harv} 
\bibliography{LOD_ref}

\end{document}